# Robust Quantum Controllers: Quantum Information - Thermodynamic Hidden Force Control in Intelligent Robotics based on Quantum Soft Computing

Ulyanov Sergey V.*, Ulyanov Viktor S.† and Hagiwara T. **

*Institute of System Analysis and Management, Dubna State University
* Meshcheryakov Laboratory of Information Technologies, Joint Institute for Nuclear Research (JINR)
†Department of Information Technologies, Moscow State University of Geodesy and Cartography (MIIGAiK)
**Yamaha Motor Co. Ltd., Automotive operations Dpt.
*Email: srg.v.ulyanov@gmail.com
†Email: ulyanovik@gmail.com
** Email: hagiwarat@yamaha-motor.co.jp

**Abstract**

A generalized strategy for the design of intelligent robust control systems based on quantum / soft computing technologies is described. The reliability of hybrid intelligent controllers increase by providing the ability to self-organize of imperfect knowledge bases. The main attention is paid to increasing the level of robustness of intelligent control systems in unpredictable control situations with the demonstration by illustrative examples. A SW & HW platform and support tools for a supercomputer accelerator for modeling quantum algorithms on a classical computer are described.

## Introduction

For complex and ill-defined dynamic control objects that are not easily controlled by conventional control systems (such as *P*-[*I*]-*D*-controllers) — especially in the presence of fuzzy model parameters and different stochastic noises — the System of Systems Engineering methodology provides fuzzy controllers (FC) as one of alternative way of control systems design.

Soft computing methodologies, such as genetic algorithms (GA) and fuzzy neural networks (FNN) had expanded application areas of FC by adding optimization, learning and adaptation features.

But still now it is difficult to design optimal and robust intelligent control system, when its operational conditions have to evolve dramatically (aging, sensor failure and so on). Such conditions could be predicted from one hand, but it is difficult to cover such situations by a single FC.

Using unconventional computational intelligence toolkit, we propose a solution of such kind of generalization problems by introducing a *self-organization* design process of robust KB-FC that supported by the *Quantum Fuzzy Inference* (QFI) based on quantum soft computing ideas [1-3].

## 1. Problem's Formulation

### A. *Main problem and toolkit*

One of main problem in modern FC design is how to design and introduce robust KBs into control system for increasing *self-learning, self-adaptation and self-organizing capabilities* that enhance robustness of developed FC in unpredicted control situations.

The *learning* and *adaptation* aspects of FC's have always the interesting topic in advanced control theory and system of systems engineering. Many learning schemes were based on the *back-propagation* (BP) algorithm and its modifications (see, for example, [3] and their references). Adaptation processes are based on iterative stochastic algorithms. These ideas are successfully working if we perform our control task without a presence of ill-defined stochastic noises in environment or without a presence of unknown noises in sensors systems and control loop, and so on.

For more complicated control situations learning and adaptation methods based on BP-algorithms or iterative stochastic algorithms do not guarantee the required robustness and accuracy of control.

The solution of this problem based on SCO was developed in [2]. For achieving of *self-organization* level in intelligent control system it is necessary to apply QFI [3, 4]. The described *self-organizing* FC design method is based on special form of QFI that uses a few of partial KBs designed by SCO.

QFI uses the laws of quantum computing technologies [5] and explores three main unitary operations: (i) superposition; (ii) entanglement (quantum correlations); and (iii) interference. According to quantum gate computation, the logical union of a few KBs in one generalized space is realized with *superposition* operator; with *entanglement* operator (that can be equivalently described by different models of *quantum oracle* [6]) a search of a «successful» marked solution is formalized; and with *interference* operator we can extract «good» solutions with classical *measurement* operations [7].

### ***B.*** *Method of solution*

Proposed QFI system consists of a few KB-FCs, each of which has prepared for appropriate conditions of control object and excitations by Soft Computing Optimizer (SCO) [2]. QFI system is a new quantum control algorithm of self-organization block, which performs post processing of the results of fuzzy inference of each independent FC and produces in on-line the generalized control signal output [4].

In this case the output of QFI is an optimal robust control signal, which combines best features of each independent FC outputs. Therefore, the operation area of such a control system can be expanded greatly as well as its robustness.

Robustness of control is the background for support the reliability of advanced control accuracy in uncertainty and information risk [5].

The simulation example of robust intelligent control based on QFI is introduced.

### ***C.*** *Main goal*

The main technical purpose of QFI is to supply a self-organization capability for many (sometimes unpredicted) control situations based on a few KBs. QFI produces robust optimal control signal for the current control situation using a reducing procedure and compression of redundant information in KB's of individual FCs. Process of rejection and compression of redundant information in KB's uses the laws of quantum information theory [5-7].

Decreasing of redundant information in KB-FC increases the robustness of control without loss of important control quality as reliability of control accuracy. As a result, a few KB-FC with QFI can be adapted to unexpected change of external environments and to uncertainty in initial information.

We introduce main ideas of quantum computation and quantum information theory [6] applied in developed QFI methods. *Quantum Fuzzy Inference* ideas are introduced. Robustness of new types of *self-organizing intelligent control systems* is demonstrated.

## 2. SCO-structure based on soft computing

### ***D.*** *KB of FC creation*

SCO uses the chain of GAs ($GA_1$, $GA_2$, $GA_3$) and approximates measured or simulated data (TS) about the modeled system with desired accuracy or using real robot for it. $GA_1$ solves optimization problem connected with the optimal choice of number of membership functions and their shapes. $GA_2$ searches optimal KB with given level of rules activation. Introduction of activation level of rules allows us to sort fuzzy rules in accordance with value information and design robust KB. $GA_3$ refines KB by using a few criteria.

*Figure 1* shows the flow chart of SCO operations on macro level and combines several stages.

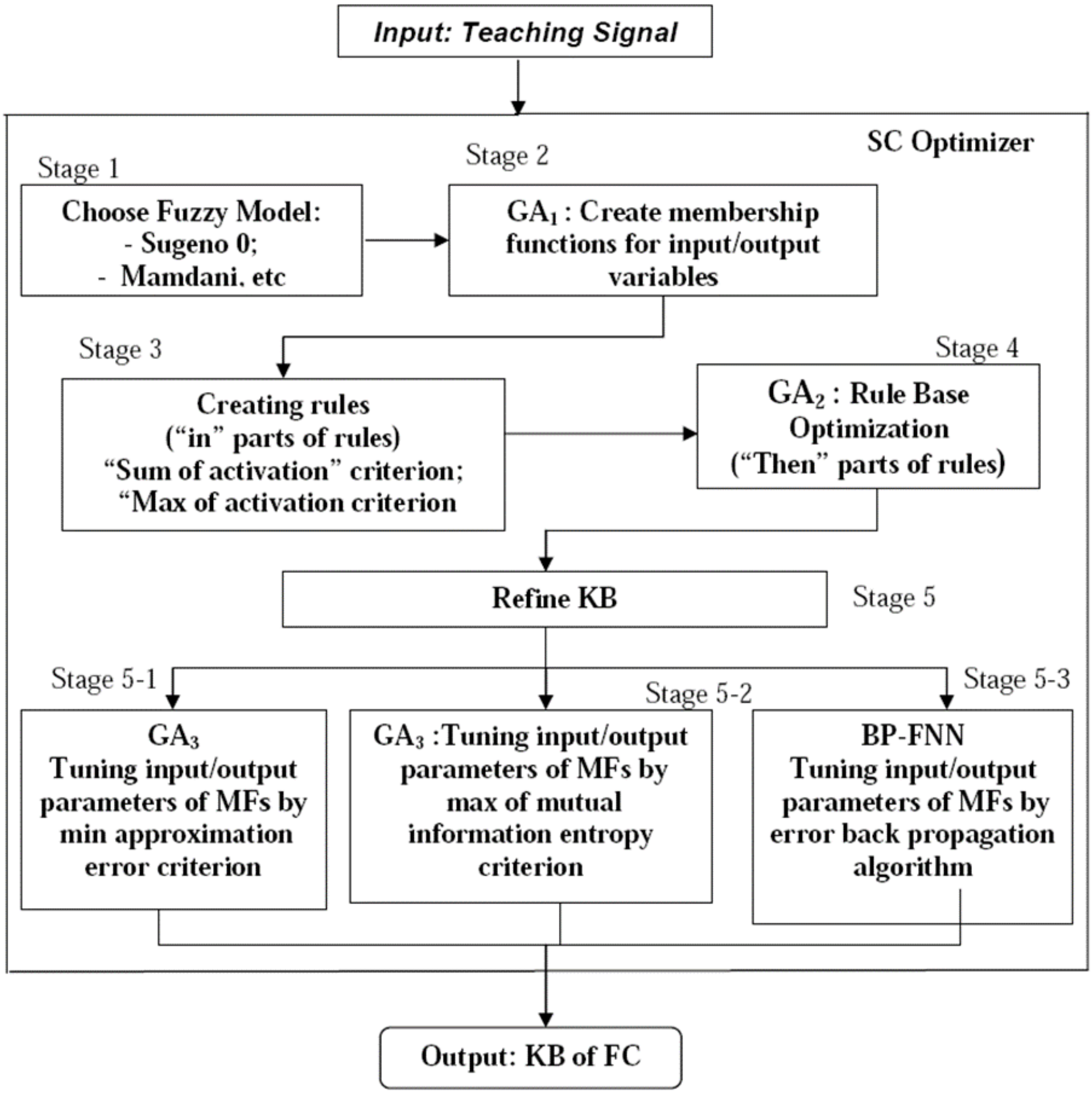

*Figure 1: Flow chart of SC Optimizer.*

Stage 1: *Fuzzy Inference System* (FIS) *Selection*. The user makes the selection of fuzzy inference model with the featuring of the following initial parameters: Number of input and output variables; Type of fuzzy inference model (Mamdani, Sugeno, Tsukamoto, etc.); Preliminary type of MFs.

Stage 2: *Creation of linguistic values*. By using the information (that was obtained on Stage 1), $GA_1$ optimizes membership functions number and their shapes, approximating teaching signal (TS), obtained from the in-out tables, or from dynamic response of control object (real or simulated in Matlab).

Stage 3: *Creation rules*. At this stage we use the rule rating algorithm for selection of certain number of selected rules prior to the selection of the index of the output membership function corresponding to the rules. For this case two criteria based on a rule's activation parameter called as a «manual threshold level» (TL). This parameter is given by a user (or it can be introduced automatically).

Stage 4: *Rule base optimization*. $GA_2$ optimizes the rule base obtained on the Stage 3, using the fuzzy model obtained on Stage 1, optimal linguistic variables, obtained on Stage 2, and the same TS as it was used on Stage 1. Rule base optimization can be performed by using mathematical model, or by using distance connection to real control object.

Stage 5: *Refine KB*. On this stage, the structure of KB is already specified and close to global optimum.

In order to reach the optimal structure, a few methods can be used. First method is based on $GA_3$ with fitness function as minimum of approximation error, and in this case KB refining is similar to classical derivative based optimization procedures (like error back propagation (BP) algorithm for FNN tuning). Second method is also based on $GA_3$ with fitness function as maximum of mutual information entropy. Third method is realized as pure error back propagation (BP) algorithm. BP algorithm may provide further improvement of output after genetic optimization. As output results of the Stages 3, 4 and 5, we have a set of KB corresponding to chosen KB optimization criteria.

***E.*** *Remote rule base optimization*

Remote KB optimization is performed on the fourth stage of designing FC (*Fig. 2*). The implementation of the physical environment connection intends to use additional equipment for the data transfer, such as radio channel, Bluetooth, WiFi or a cable connection, such as USB. Exchange of information between the management system and the SCO intended to form a KB (*Fig. 2*).

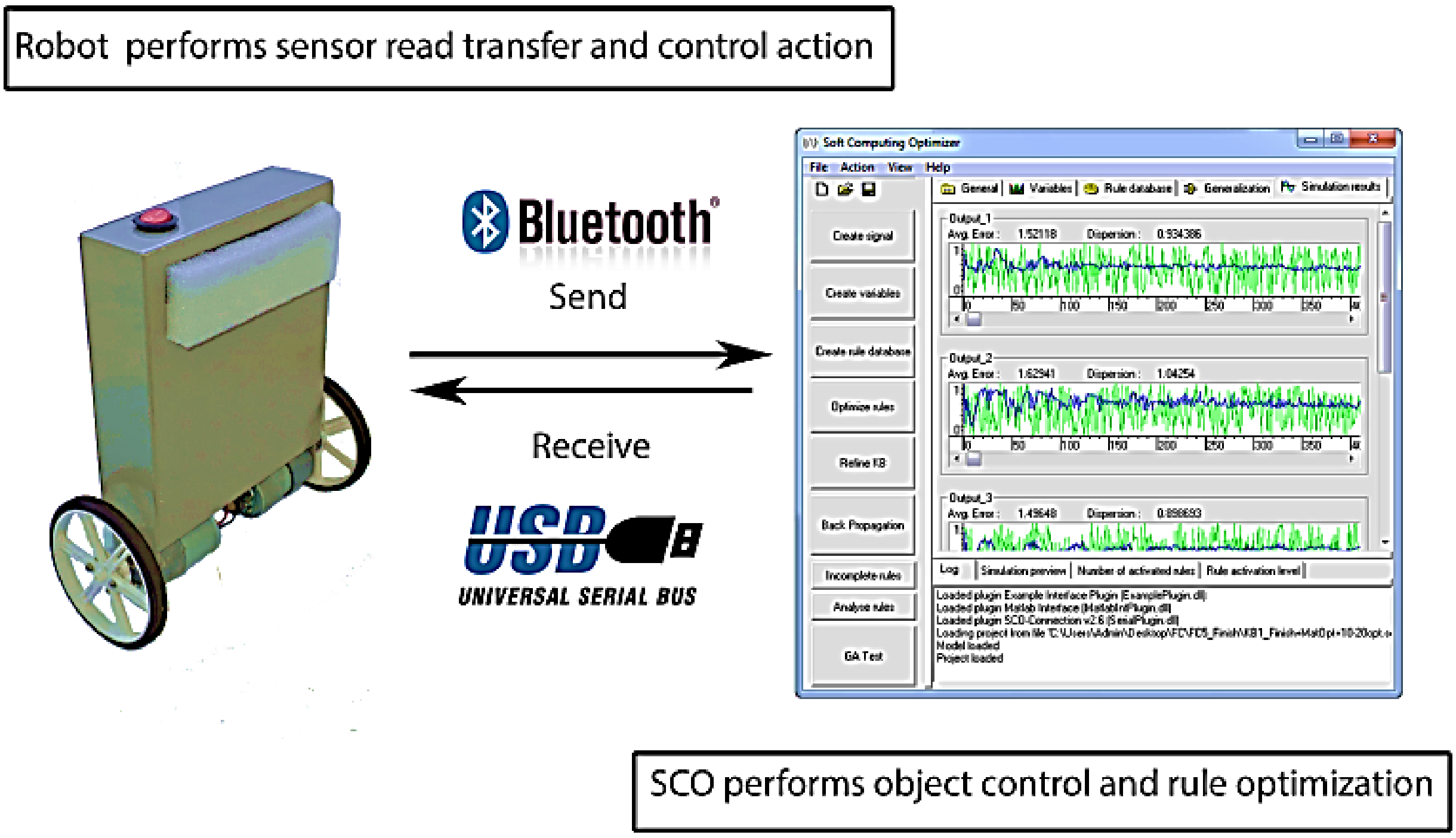

*Figure 2: Remote rule base optimization scheme.*

The control system reads the sensors and sends data to a computer for further processing. By taking input values, SCO evaluates previous decision (KB-FC) and performs fuzzy inference to check the following solutions (KB-FC). The result of the fuzzy inference is sent to the remote device. Thereafter, the control system by processing the input values generates control action.

Synchronization of SCO and control systems is based on the remote device (robot). To this end, a special program (firmware) is developed.

Connection profile uses the serial port. Transmission rate in this case is 115,200 bits / sec. During operation, floats in symbolic form are passing via COM-port. Connection to SCO uses designed plug-in. Before establishing a connection to the SCO, COM port number and the check time of one solution (the number of cycles of the system to test solution) are selected.

## 3. QFI-structure based on quantum computing

For design of QFI based on a few KBs it is needed to apply the additional operations to partial KBs outputs that drawing and aggregate the value information from different KBs. Soft computing tool does not contain corresponding necessary operations [8].

The necessary unitary reversible operations are called as *superposition, entanglement* (quantum correlation) and *interference* that physically are operators of quantum computing in information processing.

We introduce briefly the particularities of quantum computing and quantum information theory that are used in the quantum block QFI (*Fig. 3*) supporting a self-organizing capability of FC in robust intelligent control system (ICS).

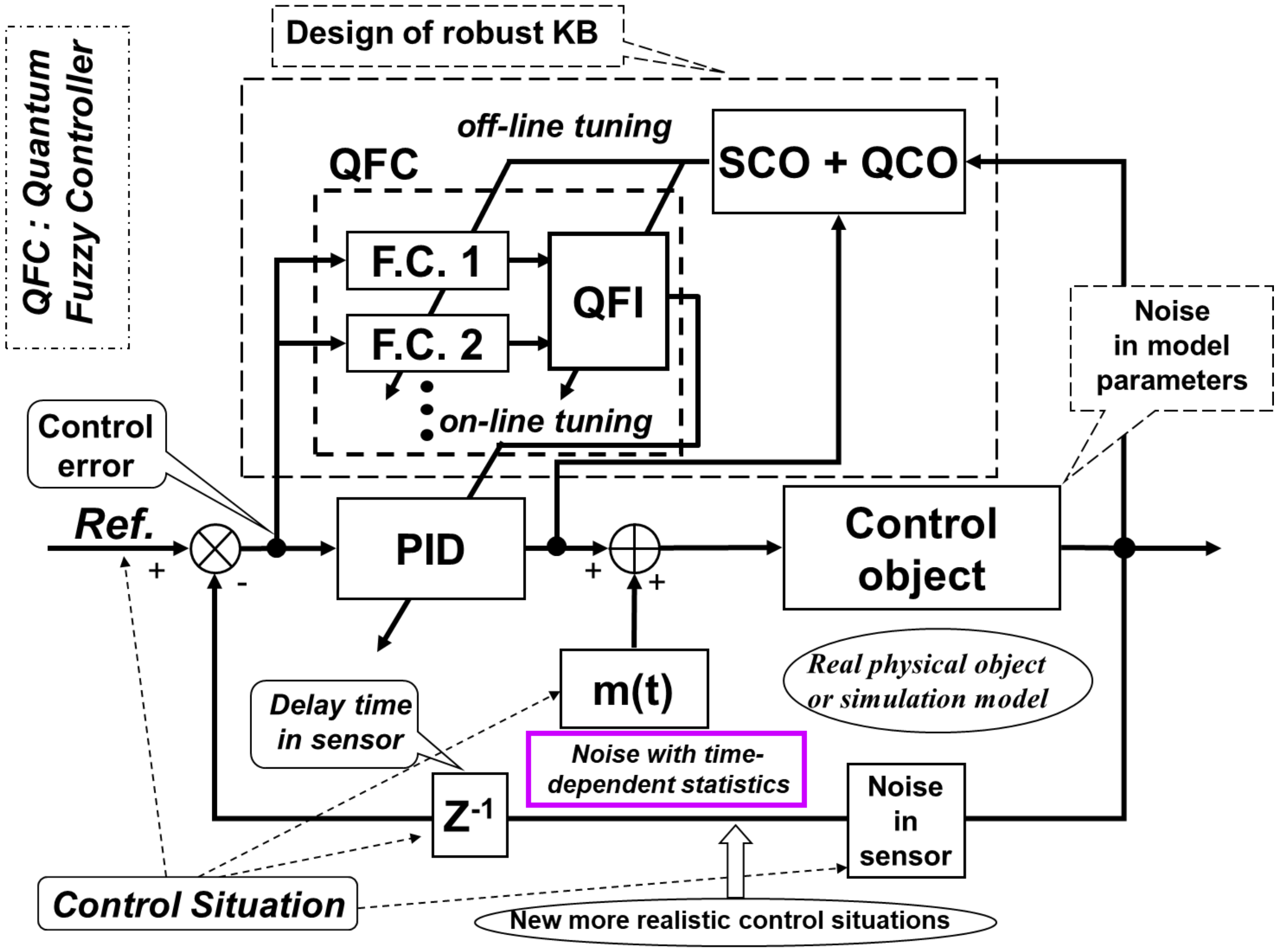

*Figure 3: Structure of robust ICS based on QFI.*

Let us consider peculiarities of quantum computing.

***F.** Quantum computing*

In Hilbert space the superposition of classical states $\left(c_1^{(1)}|0\rangle + c_2^{(1)}|1\rangle\right)$ called quantum bit (qubit) means that «*False*» and «*True*» are jointed in one state with different probability amplitudes, $c_i^1, i = 1,2$. If the Hadamard transform $H = \frac{1}{\sqrt{2}}\begin{pmatrix} 1 & 1 \\ 1 & -1 \end{pmatrix}$ is independently applied to different classical states then a tensor product of superposition states is the result:

$$|\psi\rangle = H^{\otimes n}|False\rangle = \frac{1}{\sqrt{2^n}} \otimes_{i=1}^{n} \left(|False\rangle + |True\rangle\right). \quad \textbf{(1)}$$

The fundamental result of quantum computation stays that all of the computation can be embedded in a circuit, which nodes are the universal gates. These gates offer an expansion of unitary operator $U$ that evolves the system in order to perform some computation. Thus, naturally two problems are discussed: (i) Given a set of functional points $S = \{(x, y)\}$ find the operator $U$ such that $y = U \cdot x$; (ii) Given a problem, fined the quantum circuit that solves it.

Algorithms for solving these problems may be implemented in a hardware quantum gate or in software as computer programs running on a classical computer.

It is shown that in quantum computing the construction of a universal quantum simulator based on classical effective simulation is possible [3,6,7].

In the general form, the model of quantum algorithm computing comprises the following five stages:

- preparation of the initial state $|\psi_{out}\rangle$ (classical or quantum);
- execution of the Hadamard transform for the initial state in order to prepare the superposition state;
- application of the entangled operator or the quantum correlation operator (quantum oracle) to the superposition state;
- application of the interference operator;
- application of the measurement operator to the result of quantum computing $|\psi_{out}\rangle$.

Hence, a quantum gate approach can be used in a global optimization of KB structures of ICSs that are based on quantum computing, on a quantum genetic search and quantum learning algorithms [8].

### *G. Quantum information resources in QFI algorithm*

*Figure 4* shows the algorithm for coding, searching and extracting the value information from two KBs of fuzzy PID controllers designed by SCO.

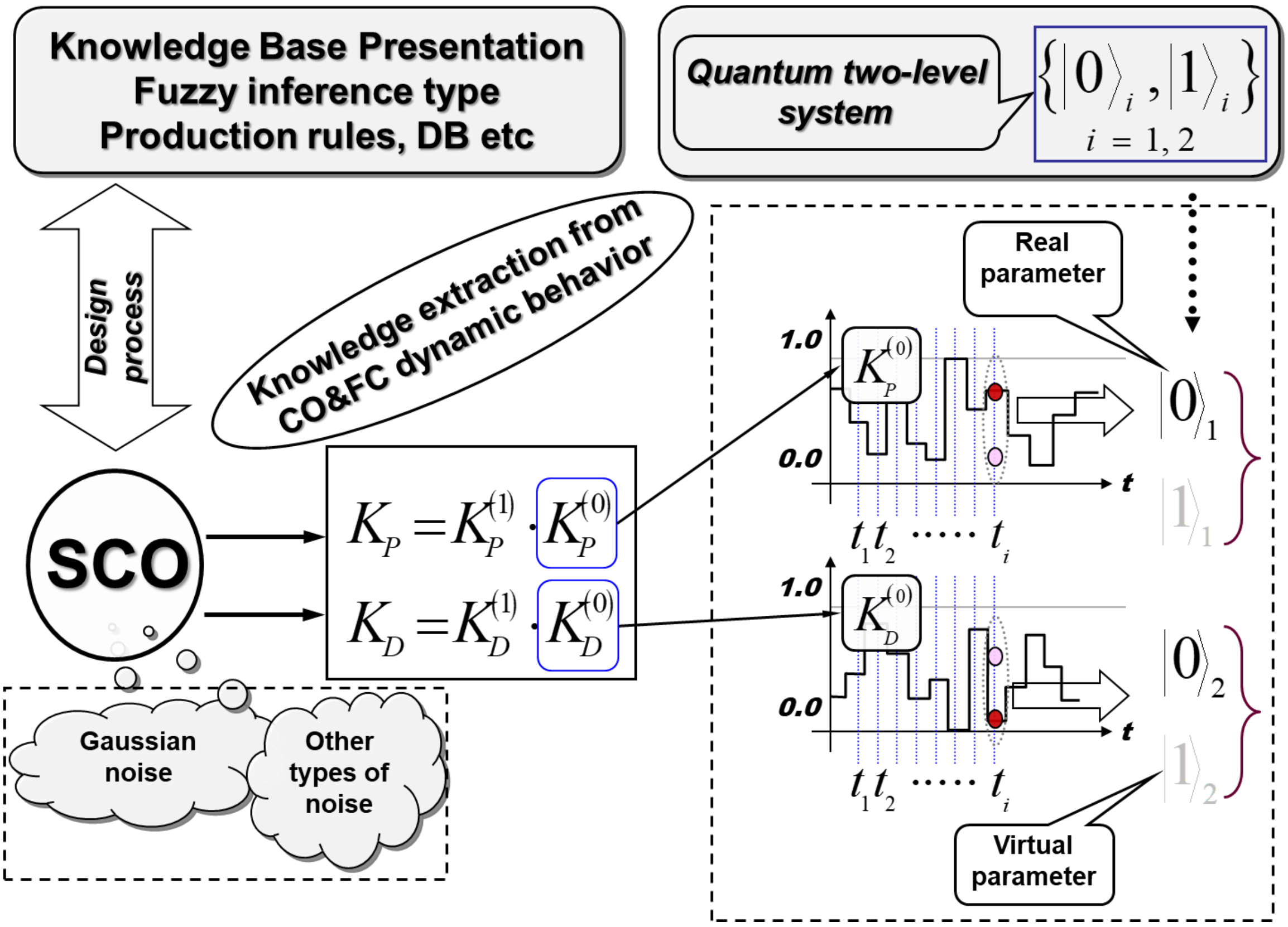

*Figure 4: Example of information extraction in QFI.*

Thus, the quantum algorithm for QFI (*Fig. 5*) the following actions are realized [5]:

- The results of fuzzy inference are processed for each independent FC;
- Based on the methods of quantum information theory, valuable quantum information hidden in independent (individual) knowledge bases is extracted;

- In on-line, the generalized output robust control signal is designed in all sets of knowledge bases of the fuzzy controller.
- In this case, the output signal of QFI in on-line is an optimal signal of control of the variation of the gains of the PID controller, which involves the necessary (best) qualitative characteristics of the output control signals of each of the fuzzy controllers, thus implementing the self-organization principle.

Therefore, the domain of efficient functioning of the structure of the intelligent control system can be essentially extended by including robustness, which is a very important characteristic of control quality.

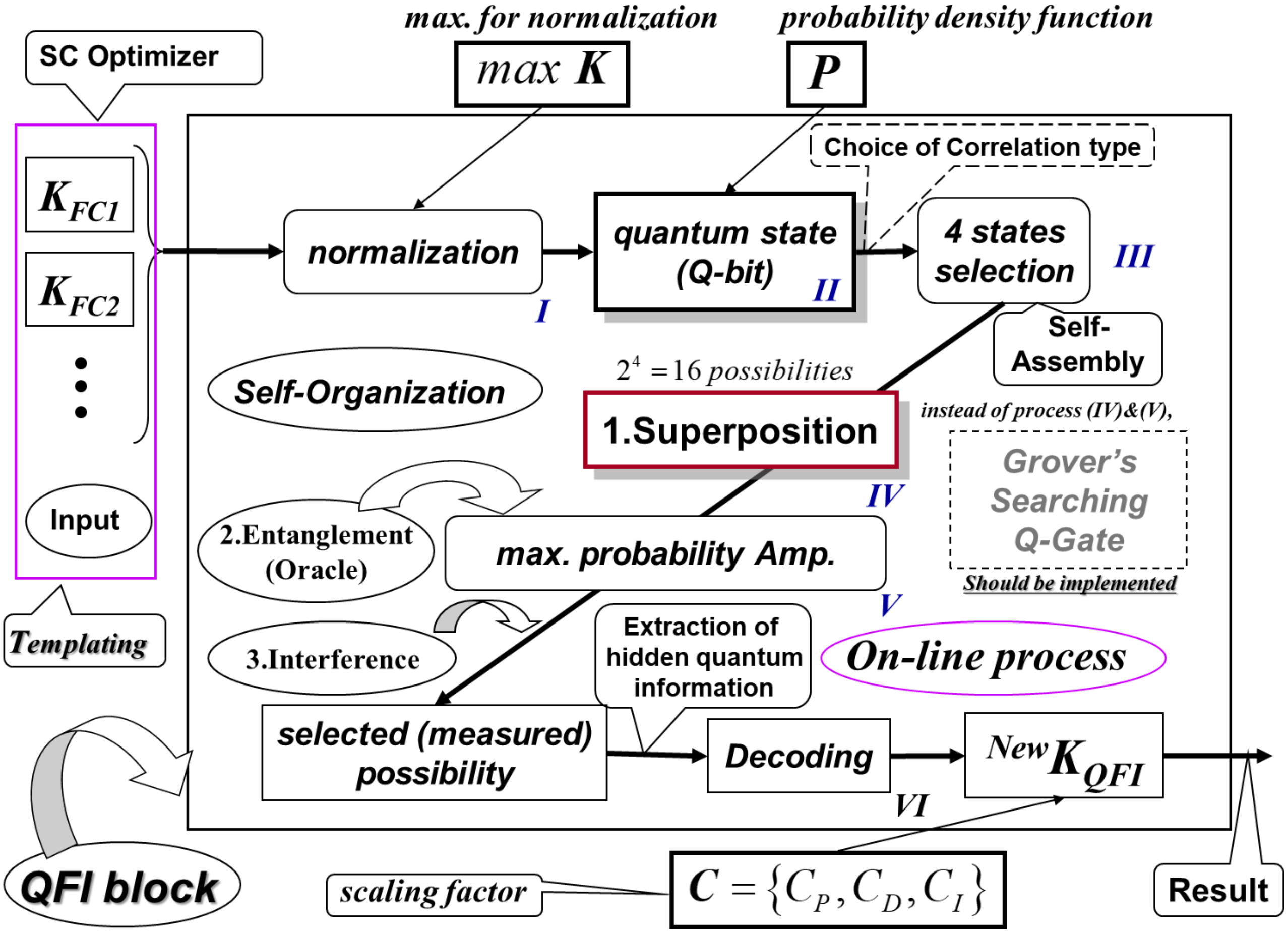

*Figure 5: The structure of QFI gate.*

The robustness of the control signal is the background for maintaining the reliability and accuracy of control under uncertainty conditions of information or a weakly formalized description of functioning conditions and/or control goals.

QFI model based on physical laws of quantum information theory, for computing use unitary invertible (quantum) operators and they have the following names: *superposition*, *quantum correlation* (entangled operators), and *interference*. The fourth operator, measurement of result quantum computation is irreversible.

Optimal drawing process of value information from a few KBs that are designed by soft computing is based on following four facts from quantum information theory [4]: (i) the effective quantum data compression; (ii) the splitting of classical and quantum parts of information in quantum state; (iii) the total correlations in quantum state are «mixture» of classical and quantum correlations; and (iv) the exiting of hidden (locking) classical correlation in quantum state [6,9].

This quantum control algorithm uses these four Facts from quantum information theory: (i) compression of classical information by coding in computational basis $\{|0\rangle,|1\rangle\}$ and forming the quantum correlation between different computational bases (Fact 1); (ii) separating and splitting total information and correlations on «classical» and «quantum» parts using Hadamard transform (Facts 2 and 3); (iii) extract unlocking information and residual redundant information by measuring the classical correlation in quantum state (Fact 4) using criteria of maximal corresponding amplitude probability.

These facts are the informational resources of QFI background. Applying these facts it is possible to extract an additional amount of quantum value information from smart KBs produced by SCO for design a *wise* control using compression and rejection procedures of the redundant information in a classical control signal. Below we discuss the application of this quantum control algorithm in QFI structure.

***H.*** *Remote quantum base optimization*

As the adjustable parameter scaling factor is used in remote quantum base optimization. Scaling factor is used in the final step of forming the gain of PID (*Fig. 5*).

During operation, floats in symbolic form are passed via COM-port. The control system reads the sensors and sends them to a computer for further processing. By taking the input values, the GA evaluates the previous decision, and carries a quantum fuzzy inference to check the following solutions. The result of the fuzzy inference is sent to the remote device. Thereafter, the control system by processing the input values generates control action. Connecting to QFI developed through a plug-in.

Before establishing a connection to the SCO, COM port number and the check time of one solution (the number of cycles of the system to test solution) are selected (*Fig. 6*).

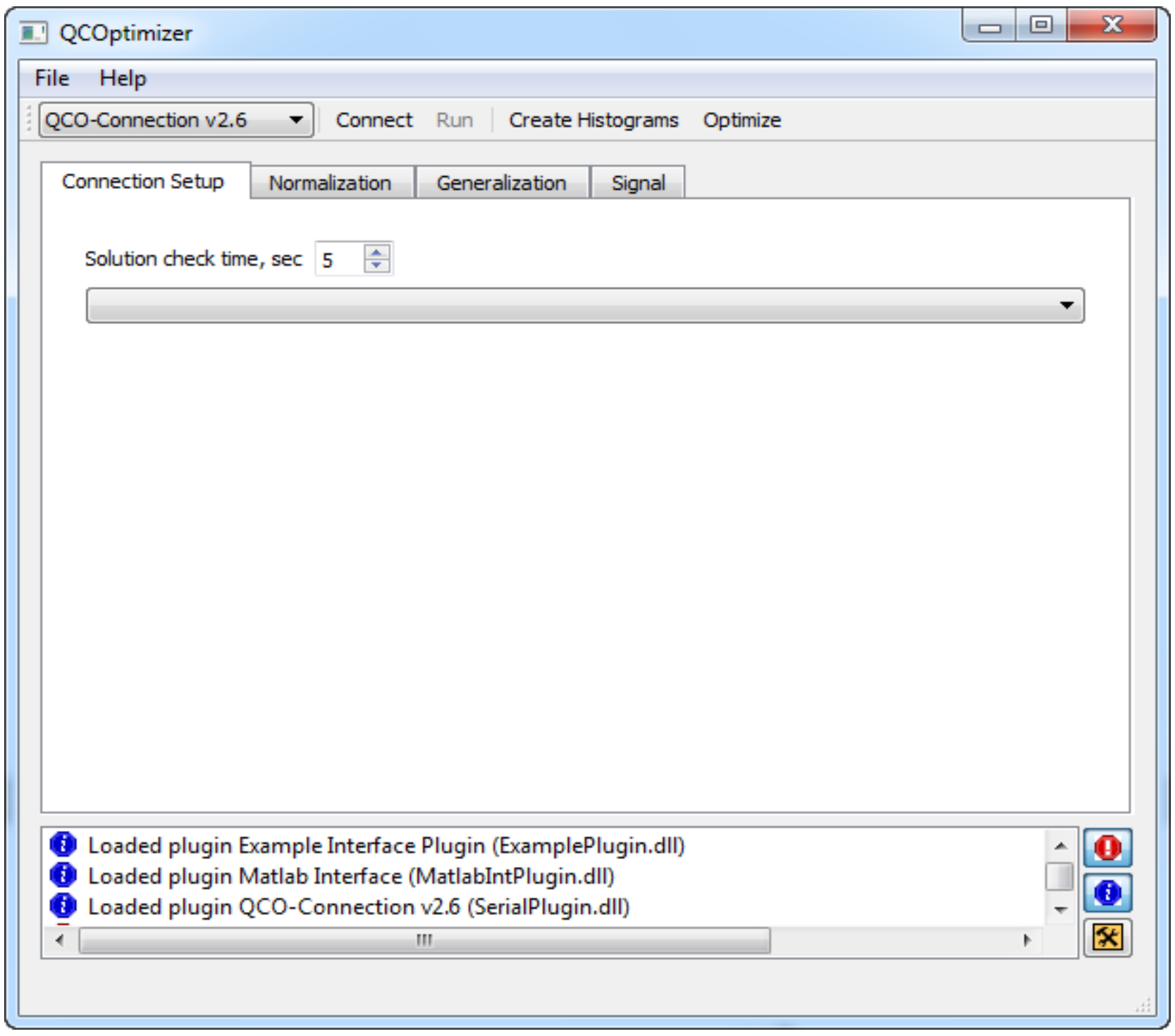

*Figure 6: Remote connection plug-in for QC Optimizer.*

## 4. KB-self-organization of FC's based on QFI

***I.*** *Robust FC design toolkit*

The kernel of the abovementioned FC design toolkit is a so-called SCO implementing advanced soft computing ideas. SCO is considered as a new flexible tool for design of optimal structure and robust

KBs of FC based on a chain of genetic algorithms (GAs) with information-thermodynamic criteria for KB optimization and advanced error back-propagation algorithm for KB refinement [2]. Input to SCO can be some measured or simulated data (called as «teaching signal» (TS)) about the modelling system. For TS design (or for GA fitness evaluation) we are used stochastic simulation system based on the control object model. More detail description of SCO is given in [1,2]. Below we discuss the application of this algorithm in QFI structure.

*Figure 3* illustrates as an example the structure and main ideas of self-organized control system consisting of two FC's coupling in one QFI chain that supplies a self-organizing capability. According to described above algorithm the input to the QFI gate is considered according (1) as a superposed quantum state $K_1(t) \otimes K_2(t)$, where $K_{1,2}(t)$ are the outputs from fuzzy controllers FC1 and FC2 designed by SCO (see, *Fig. 4*) for the given control task in different control situations (for example, in the presence of different stochastic noises).

The algorithm of superposition calculation is presented in *Fig. 7* and described in details in [4,5].

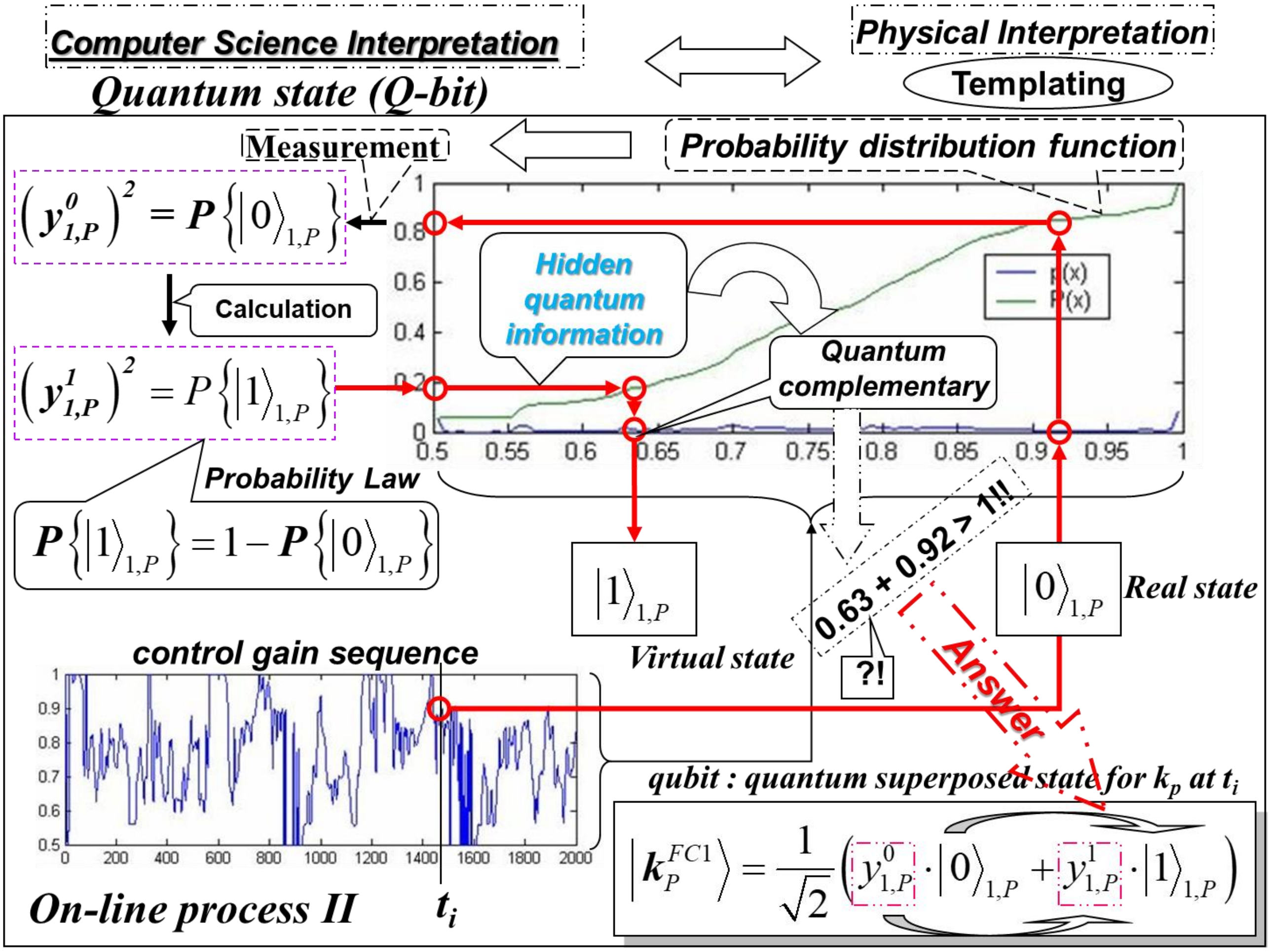

*Figure 7:* *The algorithm of superposition calculation.*

We discuss for simplicity the situation in which an arbitrary amount of correlation is unlocked with a one-way message. Let us consider the communication process between two KBs as communication between two players *A* and *B* (see, *Figs 4* and *7*) and let $d = 2^n$. According to the law of quantum mechanics, initially we must prepare a quantum state description by density matrix $\rho$ from two classical states ($KB_1$ and $KB_2$).

The initial state $\rho$ is shared between subsystems held by *A* ($KB_1$) and *B* ($KB_2$), with respective dimensions $d$ ,

$$\rho = \frac{1}{2d}\sum_{k=0}^{d-1}\sum_{t=0}^{1}\left(|k\rangle\langle k|\otimes|t\rangle\langle t|\right)_A \otimes \left(U_t|k\rangle\langle k|U_t^{\dagger}\right)_B . \qquad \textbf{(2)}$$

Here $U_0 = I$ and $U_1$ changes the computational basis to a conjugate basis $\left|\langle i|U_1|k\rangle\right| = 1/\sqrt{d} \quad \forall\, i,k$ .

In this case, $B$ chooses $|k\rangle$ randomly from $d$ states in two possible random bases, while $A$ has complete knowledge on his state. The state (2) can arise from following scenario. $A$ picks a random $n$-bit string $k$ and sends $B$ $|k\rangle$ or $H^{\otimes n}|k\rangle$ depending on whether the random bit $t = 0$ or $1$. $A$ can send $t$ to $B$ to unlock the correlation later. Experimentally, Hadamard transform, $H$ and measurement on single qubits are sufficient to prepare the state (2), and later extract the unlocked correlation in $\rho'$. The initial correlation is small, i.e. $I_{Cl}^{(l)}(\rho) = \frac{1}{2}\log d$ . The final amount of information after the complete measurement $M_A$ in one-way communication is ad hoc, $I_{Cl}(\rho') = I_{Cl}^{(l)}(\rho) = \log d + 1$, i.e., the amount of *accessible information increase*. This phenomenon is impossible classically.

However, states exhibiting this behaviour *need not be entangled* and corresponding communication can be organized using Hadamard transform [9].

Therefore, using the Hadamard transformation and a new type of quantum correlation as the communication between a few KB's it is possible to increase initial information by unconventional quantum correlation (as the quantum cognitive process of a value hidden information extraction in on-line, see, e.g. *Fig. 4*). *Figure 8* shows the structure of Quantum Computing Optimizer of robust KB-FC based on QFI [4].

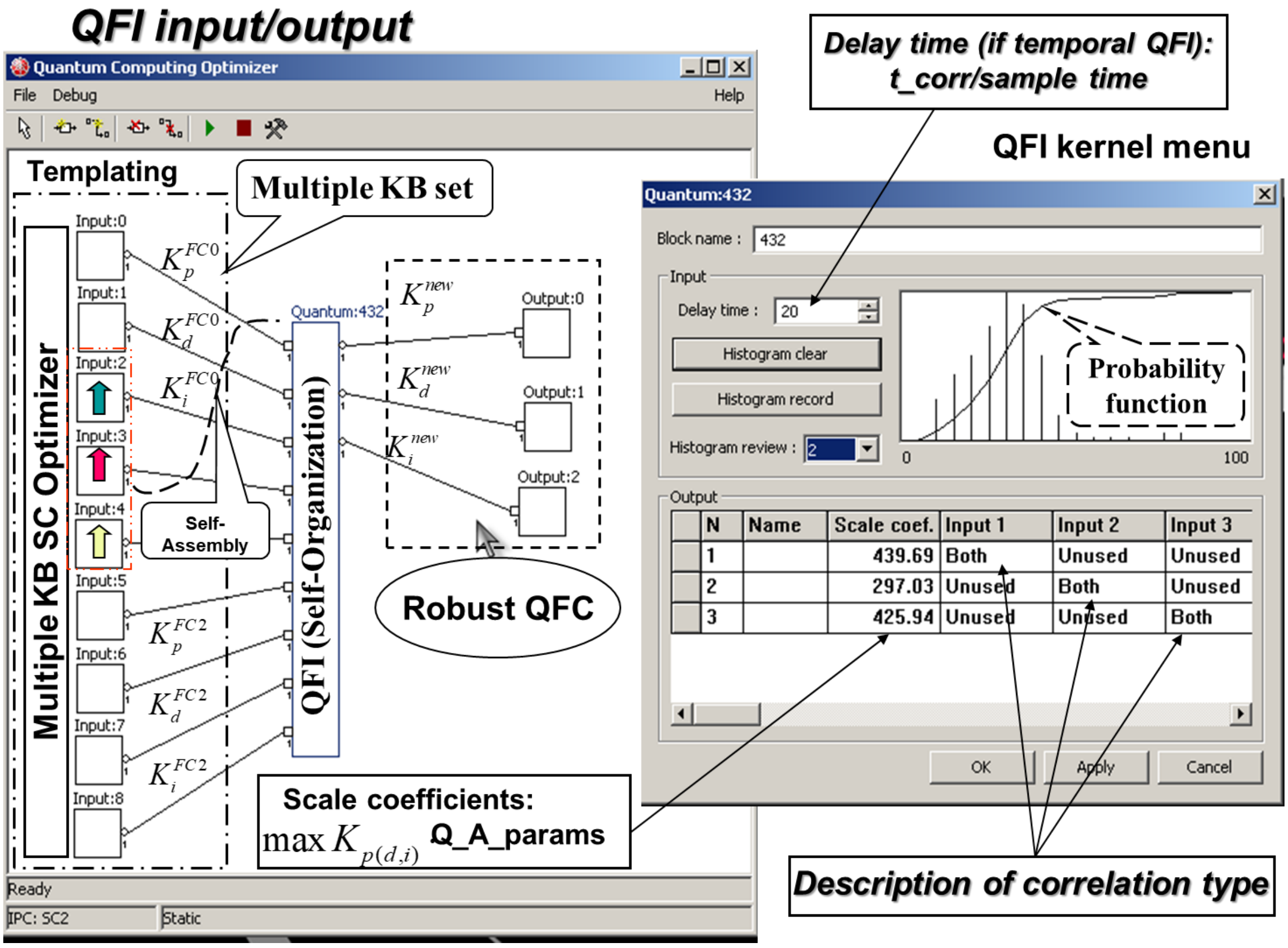

*Figure 8: QFI-process by using QC Optimizer (QFI kernel).*

In present report we consider a simplified case of QFI when with the Hadamard transform is organized an unlocked correlation in superposition of two KBs; instead of the difficult defined entanglement operation an equivalent quantum oracle is modelled that can estimates an «*intelligent*

*state*» with the maximum of amplitude probability in corresponding superposition of classical states (minimum entropy principle relative to extracted quantum knowledge [5]).

Interference operator extracts this maximum of amplitude probability with a classical measurement.

Using of described QFI model to control of non-linear locally and globally unstable dynamic systems below is described.

# 5. Benchmark's simulation

It is demonstrated that FCs prepared to maintain control object in the prescribed conditions are often fail to control when such conditions are dramatically changed. We propose the solution of such kind of problems by introducing a quantum generalization of strategies in fuzzy inference in on-line from a set of pre-defined fuzzy controllers by new QFI based systems. The latter is a new quantum algorithm in quantum computation without entanglement. Two Benchmarks are considered: robust control of locally and globally unstable control objects.

### *J. Benchmark 1: Globally unstable control object simulation*

«Cart-pole» control object is a non-linear dissipative system. This is a typical task of control theory, they demonstrating quality of control system. Task of control is the stability of inverted pendulum in vertical position. The motion of the dynamic system «cart-pole» is described by the following equations

$$\ddot{\theta} = \frac{g\sin\theta + \cos\theta\left(\dfrac{u+\xi(t)+a_1\dot{z}+a_3 z - ml\dot{\theta}^2\sin\theta}{m_c+m}\right) - k\dot{\theta}}{l\left(\dfrac{4}{3}-\dfrac{m\cos^2\theta}{m_c+m}\right)}, \qquad (3)$$

$$\ddot{z} = \frac{u+\xi(t)-a_1\dot{z}-a_2 z+ml(\dot{\theta}^2\sin\theta-\ddot{\theta}\cos\theta)}{m_c+m}, \qquad (4)$$

where $\theta$ is the pendulum deviation angle (degrees); $z$ is the movement of the cart (m); $g$ is the acceleration of gravity (9.8 m/s$^2$); $m_c$ is the pendulum mass (kg); $l$ is the pendulum half-length (m); $\xi(t)$ is the stochastic excitation; and $u$ is the control force acting on the cart (N). The equations for the entropy production rate in the control object and the PID controller have the following form, respectively:

$$\frac{d}{dt}S_\theta = \frac{k\dot{\theta}^2 + \dfrac{ml\dot{\theta}^3\sin 2\theta}{m_c+m}}{l(\dfrac{4}{3}-\dfrac{m\cos^2\theta}{m_c+m})};\ \frac{d}{dt}S_z = a_1\dot{z}^2;\ \frac{d}{dt}S_u = k_d\dot{e}^2 \qquad (5)$$

The following parameter values are determined: $m_c = 1; m = 0.1; l = 0.54 k = 0.4; a_1 = 0.1; a_2 = 5$; and the initial position $\left[\theta_0;\dot{\theta}_0;z_0;\dot{z}_0\right]=\left[10;0.1;0;0\right]$ (the value of the pendulum deviation angle is given in degrees); the constraint on the control force is –0.5 < $u$ < 5.0.

The specific feature of control problem for the given control object (4) is the application of one fuzzy PID controller for controlling the movement of the cart (with one degree of freedom), while the control object has two degrees of freedom.

The control goal is that the pendulum deviation angle (second generalized coordinate) reaches the given value via the implicit control using the other generalized coordinate and corresponding essentially nonlinear cross-connections with the cart movement coordinate (effect of energy transmission between the generalized coordinates).

***Remark 1***: *Stability Lemma for Nonlinear Systems*. Based on the relationship between thermodynamic exergy and Hamiltonian systems a fundamental stability Lemma for Hamiltonian systems formulated. The stability of Hamiltonian systems is bounded between Lyapunov and Chetaev theorems as following: Given the Lyapunov derivative as a decomposition and sum of exergy generation rate $\dot{W}$ and exergy dissipation rate $T_0\dot{S}_i$ then [10]

$$\dot{V} = \dot{W} - T_0\dot{S}_i = \sum_{j=1}^{N} Q_j\dot{q}_j - \sum_{l=1}^{M-N} Q_l\dot{q}_l \,. \tag{6}$$

where $Q_j$ is the generalized force vector and the irreversible entropy production rate can be expressed as

$$\dot{S}_i = \sum_k \mathcal{F}_k \mathcal{X}_k = \frac{1}{T_0}\sum_k Q_k\dot{q}_k \geq 0 \,.$$

A control law is Lyapunov optimal if it minimizes the first time derivative of the Lyapunov function over a space of admissible force controls. In general, a set of feedback gains are optimized by minimizing the regulating and / or tracking error of the conventional feedback controller while regulating to zero and / or tracking a desired reference input. The Lyapunov function is the total error energy which for most mechanical systems is equivalent to an appropriate Hamiltonian function H, as following: $V = \mathrm{H}$. Then the concept of Lyapunov Optimal follows directly from setting $\dot{W} = 0$ in (6) and maximizing $T_0\dot{S}_i$ for which the time derivative of the Lyapunov function (Hamiltonian) or the modified power (work / energy) equation is written as following:

$$\dot{V} = \dot{\mathrm{H}} = -T_0\dot{S}_i = -\sum_{j=1}^{N} Q_j\dot{q}_j = -\sum_{j=1}^{N} \mathcal{F}_j\dot{R}_j \,,$$

which is independent of system dynamics and is a kinematic quantity that applies to any system. Note that $F_j$ denotes a set of forces acting on a mechanical system and $\dot{R}_j$ denotes the inertial linear velocity of the point where $\mathcal{F}_j$ and in (5) is applied. Passivity control for robotic systems follows directly from setting $\dot{W} = 0$ in (6).

***Remark* 2:** *Information-like Lyapunov functions*. Recently, presented a rich information-like family of universal Lyapunov functions for any linear or non-linear reaction network with detailed or complex balance. Moreover, $H_f$ are not just Lyapunov functions but information measure of the divergences: $H_f\left(c^1(t)\middle|c^2(t)\right)$ is monotonically non-increasing function of time *t* for any two kinetic curves $c^1(t)$ and $c^2(t)$ with the same value of $\sum_i c_i$. These new functions aimed to resolve "the mystery" about the difference between the rich family of Lyapunov functions (*f* - divergences) for linear kinetics and a limited collection of Lyapunov functions for non-linear networks in thermodynamic conditions [11].

In the case of the similar initial learning conditions, the SCO with soft computing is used to design $KB_1$ of $FC_1$ for the generalized criterion of minimal mean square error:

$$\int_{t_0}^{t_{end}} \theta^2(t)\,dt + \int_{t_0}^{t_{end}} \dot{\theta}^2(t)\,dt$$

and $KB_2$ for $FC_2$ for the generalized criterion of minimal absolute error of the pendulum position:

$$\int_{t_0}^{t_{end}} \left|\ \theta(\tau)\ \right| d\tau + \int_{t_0}^{t_{end}} \left|\ \dot{\theta}(\tau)\ \right| d\tau \,.$$

Thus, we consider the solution of the vector (multi-objective) optimization problem based on the decomposition of the KB. The Gaussian noise was used as the random signal for designing $KB_1$, and Rayleigh noise was used for forming $KB_2$ (see *Fig. 9*, learning situations (**S1**, **S2**), respectively).

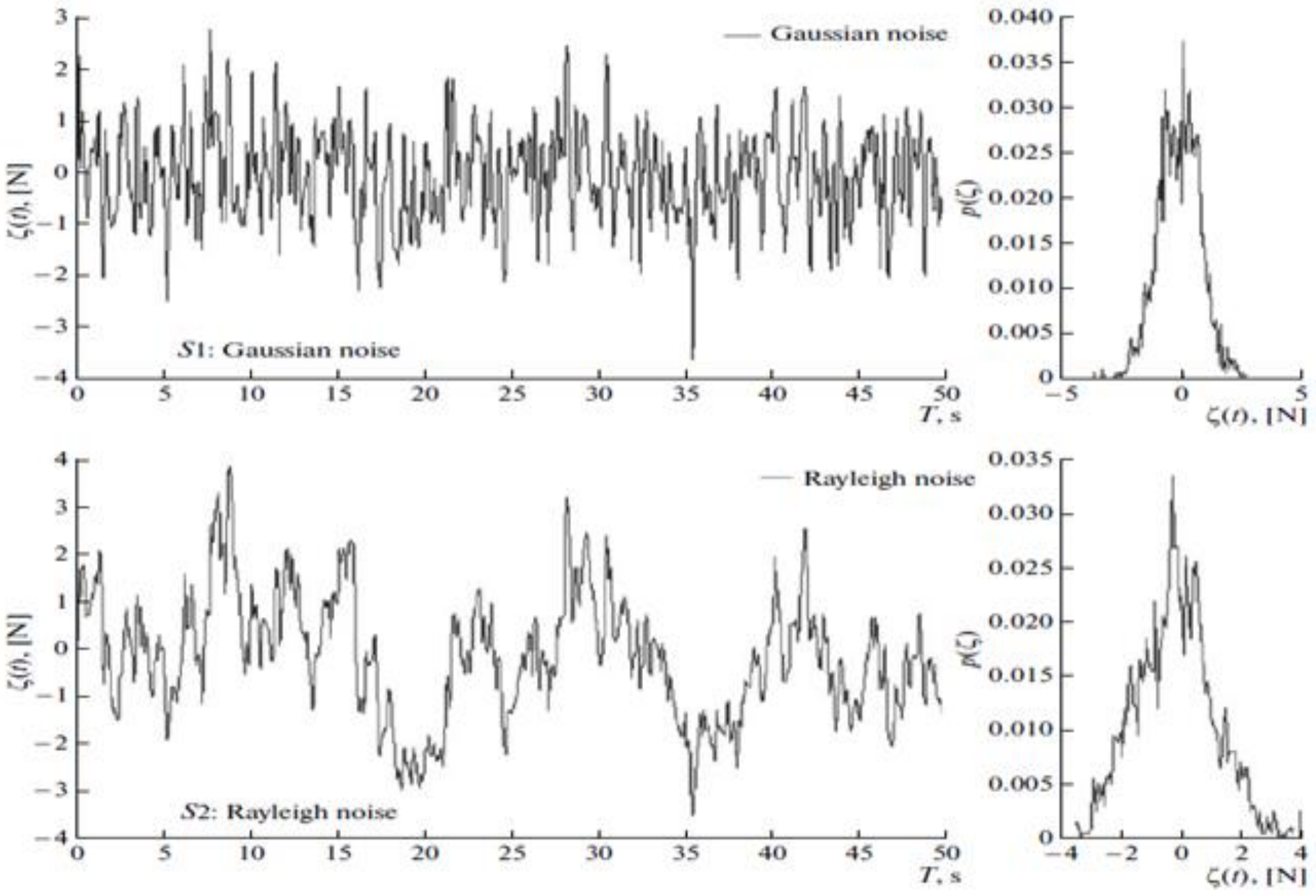

*Figure 9: Random noise used in situations (S1, S2).*

Physically the first criterion is equivalent to the total energy of the overturned pendulum and the second criterion characterizes the precision of the dynamic behavior of the control object.

*Figure 10* shows $KB_1$ and $KB_2$ with the corresponding activated numbers of rules equal to 22 and 33 for a total number of rules of 729.

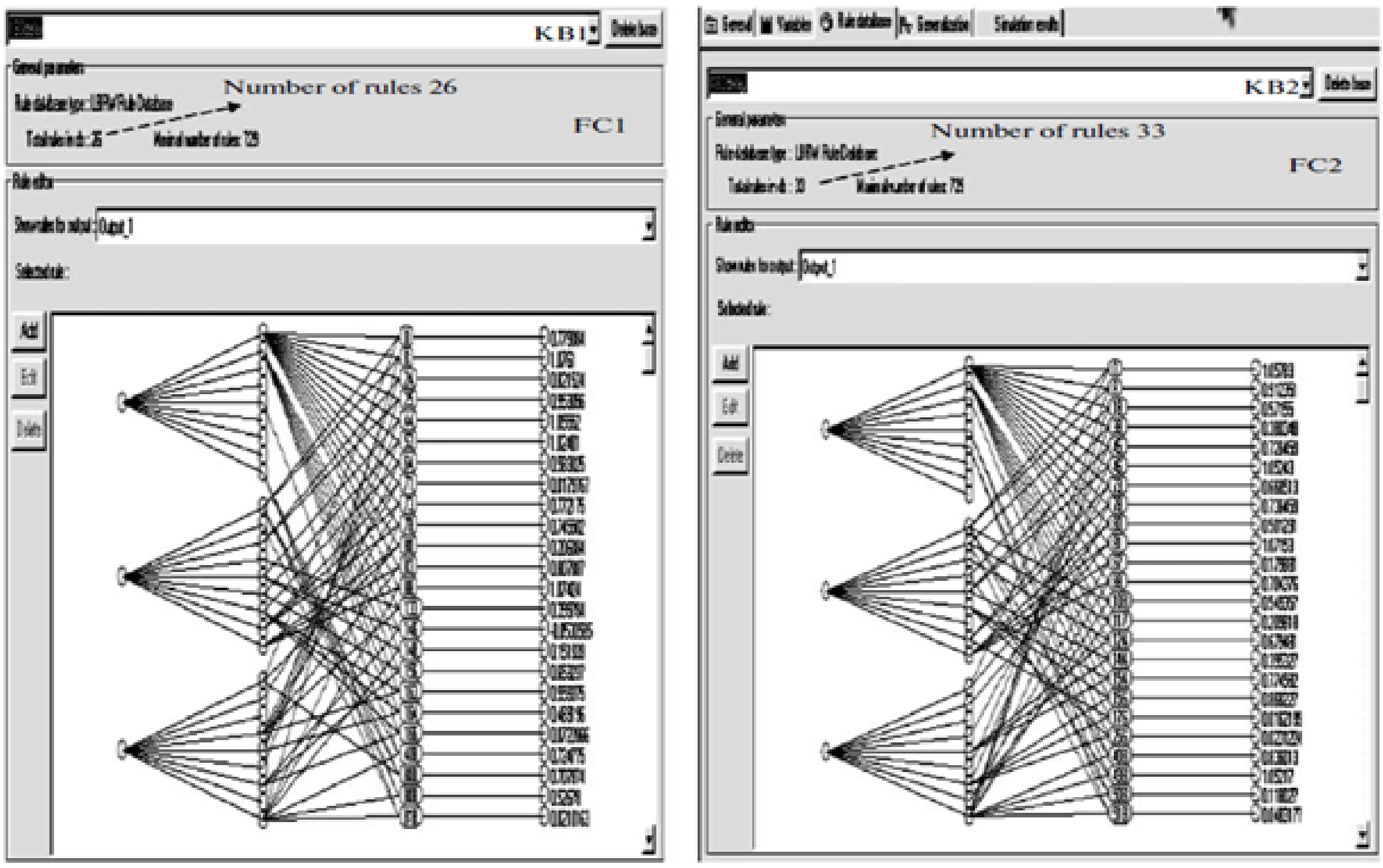

*Figure 10: Form of KB1 and KB2 with corresponding activated production rules.*

Two contingency control situations (**S3**, **S4**) were simulated; in one of them (**S3**) the new noise $\xi(t)$ was introduced, the random signal with uniform one dimensional distribution, the control error signal delay (0.03), and the noise signal in the position sensor of the pendulum (noise amplification coefficient 0.015).

*Figure 11* shows the example of operation of the quantum FC for formation of the robust control signal using the proportional gain in contingency control situation **S3**. In this case, the output signals of $KB_1$ and $KB_2$ in the form of the response on the new control error in situation **S3** are received in the quantum FC. The output of the block of quantum FC is the new signal for on line control of the factor $k_p$.

Thus, the blocks of $KB_1$ and $KB_2$, and quantum FC in *Fig. 3* form the block of KB self-organization in the contingency control situation.

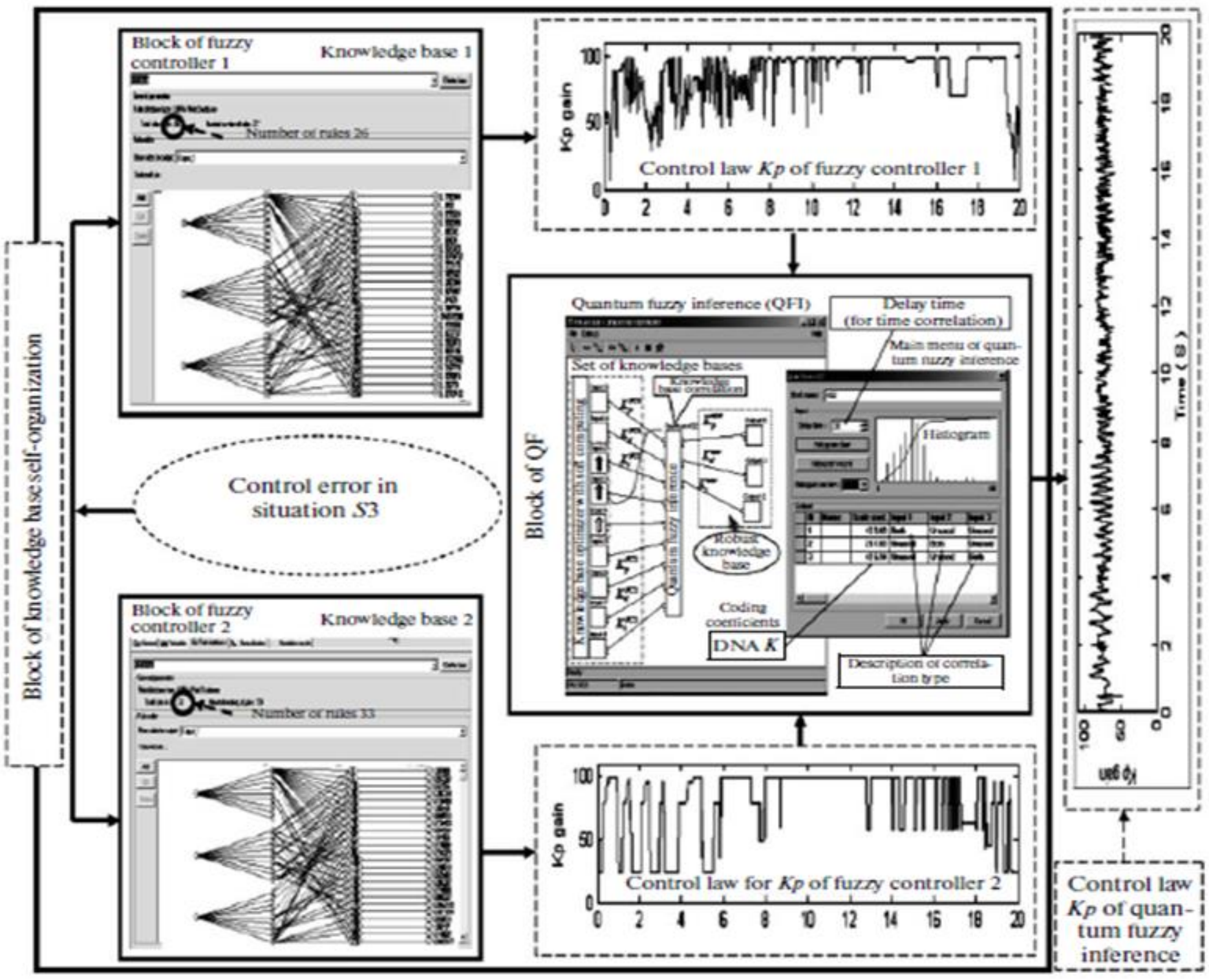

*Figure 11: Example of operation of the block of KB self-organization based on QFI.*

*Figure 12* shows the dynamic behavior of the studied system «cart-pole» and the control laws of the self-organized quantum controller (QFI), $FC_1$ and $FC_2$.

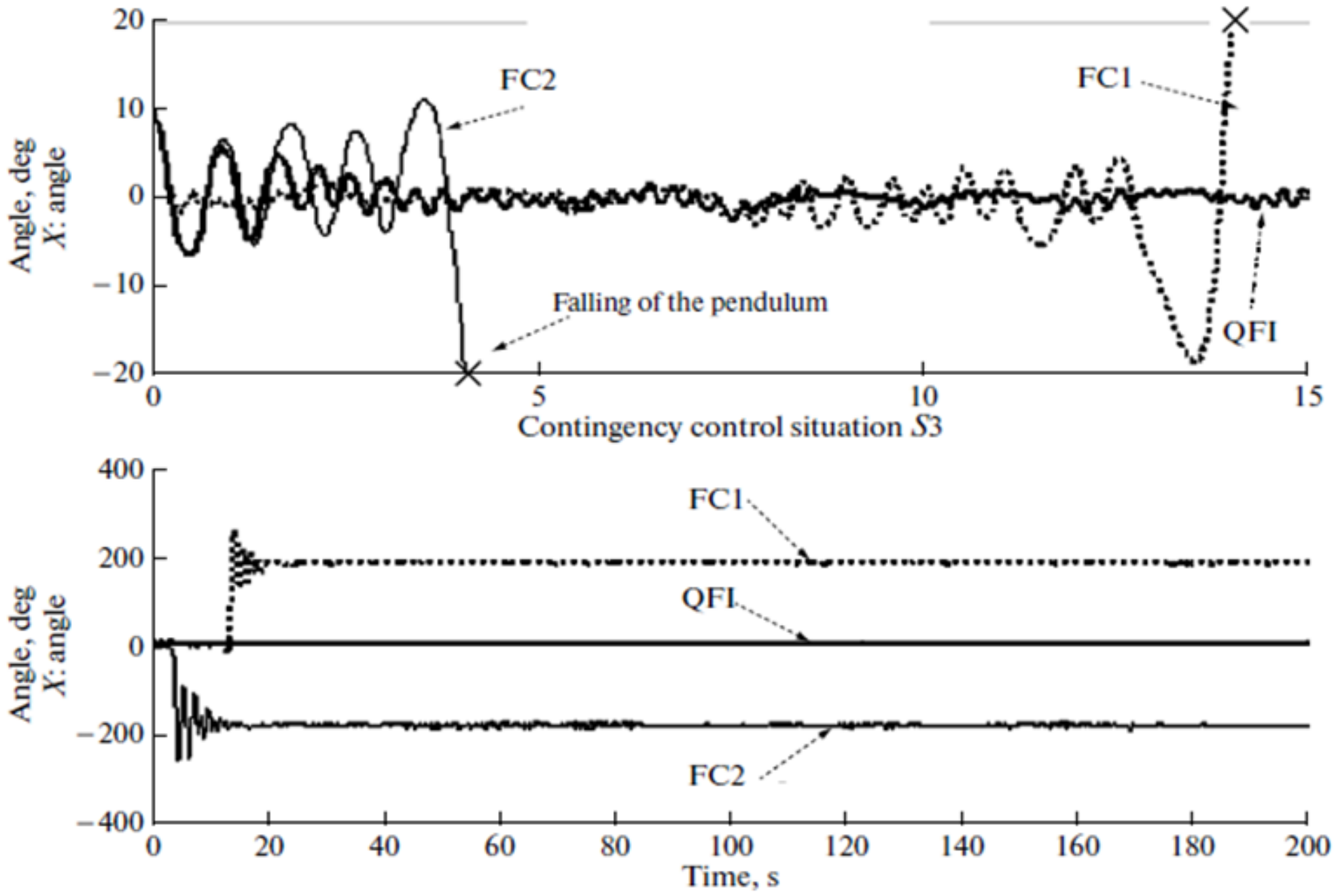

*Figure 12: Dynamic motion of pole in situation S3.*

***Remark* 3:** The following notation is used in *Fig. 12* and below: $x = \theta$ is the angle of pendulum deviation from the given position; *z* is the cart position; the quantum FC is based on the spatial correlation.

The results of simulation (*Fig. 12*) demonstrate that the dynamic control object in contingency control situations (**S3**) for the control of $FC_1$ ($FC_2$) loses stability, and for the control of quantum FC the control system possesses the property of robustness and achieving the control goal is guaranteed. According to the results of simulation (*Fig. 12*), the required amount of control for the given criteria in contingency control situations (**S3**) for the control of $FC_1$ and $FC_2$ also is not achieved, while in the case of control of the quantum FC the control system possesses the required amount of control. This yields that two non-robust fuzzy controllers can be used to design in on line the robust fuzzy controller using quantum self-organization; the KB of this robust FC satisfies both quality criteria.

Therefore, the decomposition of the solution to the above multi-objective optimization problem for the robust KB in the contingency control situation into partial solutions to optimization sub-problems physically can be performed in on line in the form of separate responses of the corresponding individual KBs optimized with different fixed cost functions and control situations.

The aggregation of the obtained partial solutions in the form of the new robust KB is performed based on the quantum FC containing the mechanism of formation of the quantum correlation between the obtained partial solutions.

As a result, only responses of the finite number of individual KBs containing limiting admissible control laws in the given contingency situations are used.

The control laws of variation of the gains of the fuzzy PID controller formed by the new robust KB have a simpler physical realization, and as a result they possess better characteristics of individual control cost function for the contingency control situation.

For experimental testing a physical model of robot (*Fig. 13*) is used.

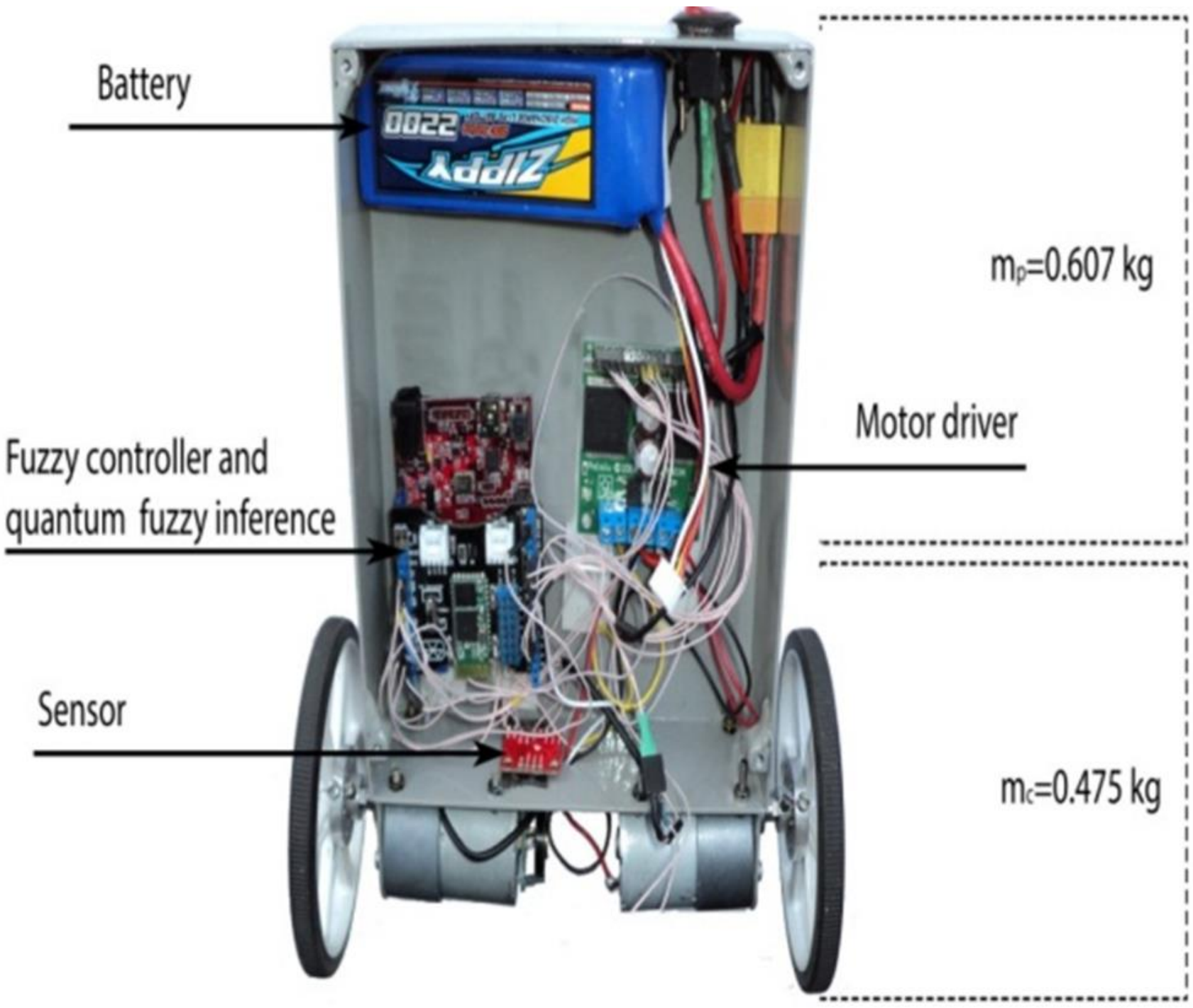

*Figure 13: Mobile robot configuration.*

Three situations of control are tested. First situation images simple situation.

The second situation use uniform noise in control channel, Gaussian noise in wheel friction and delay of control action — 0.01 s.

And the third situation have delay of control action equal 0.03s. Simulation and experimental results (for the complex situation 3) are shown on *Fig. 14*.

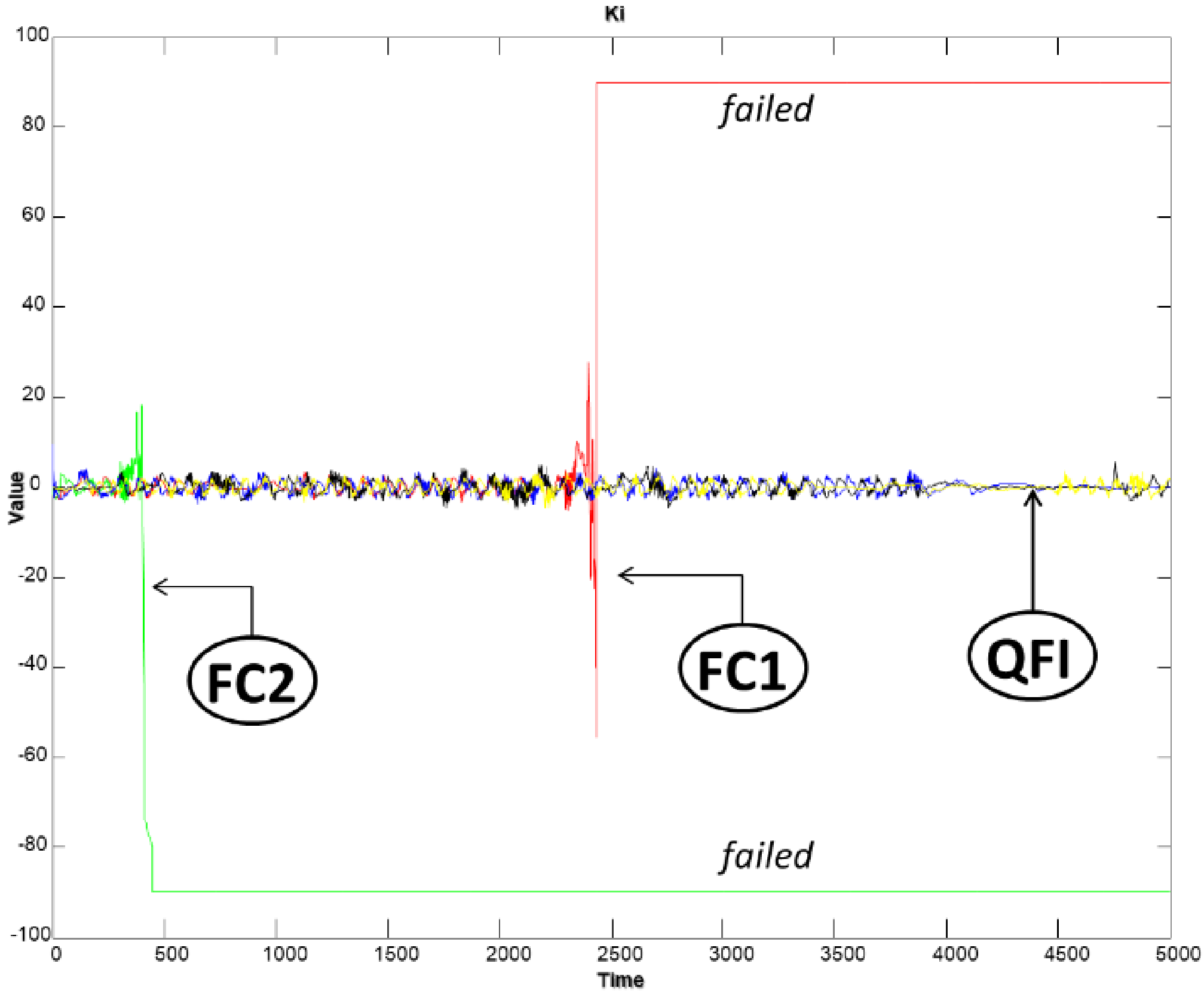

**(a)**

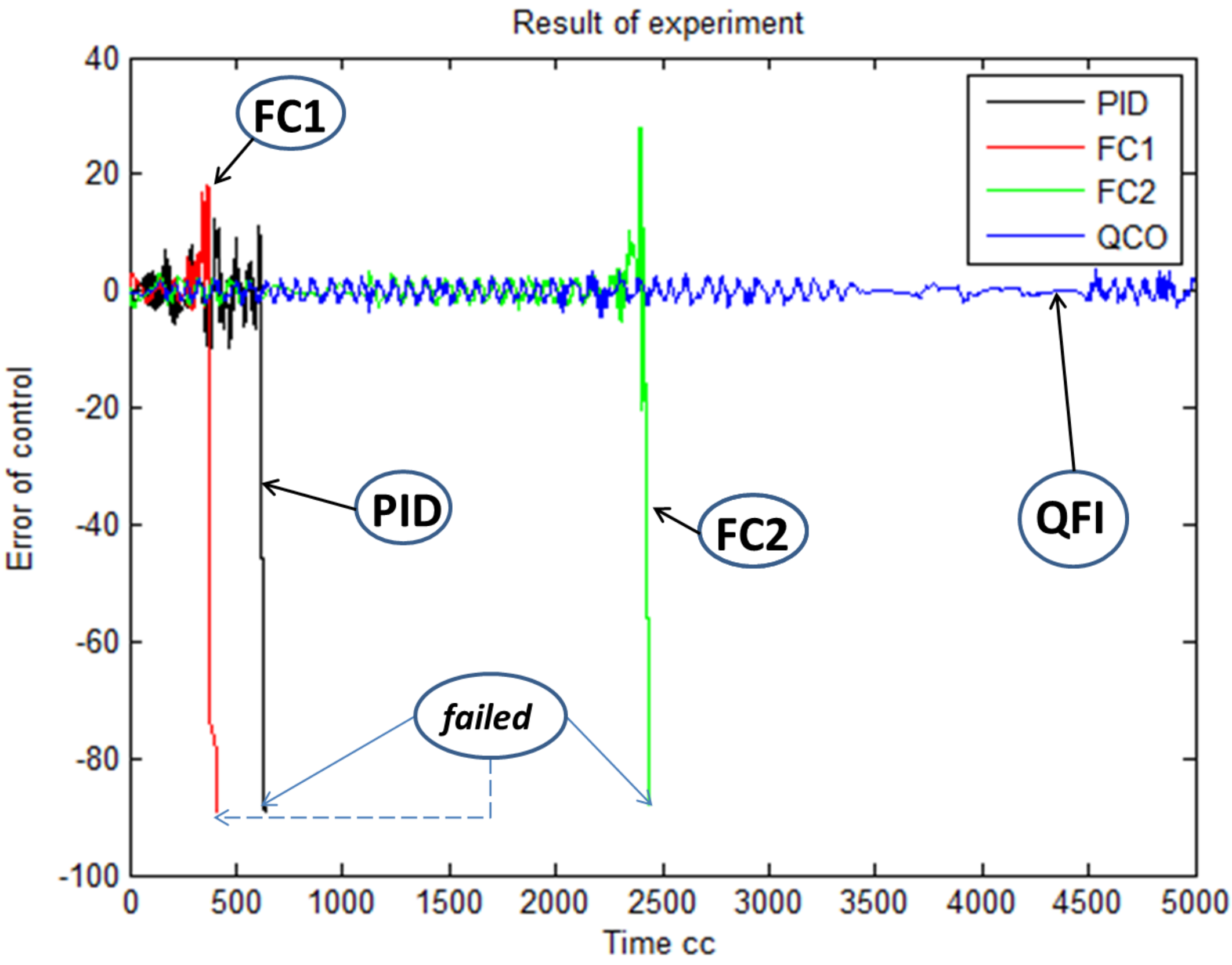

**(b)**

*Figure 14: Control error. Unpredicted situation: (a) modeling; (b) experiment on physical model.*

***K.*** *Benchmark 2: Remote rule base optimization*

To compare method of remote rule optimization on the real control object with method using Matlab simulation for optimization we created 6 KB-FC.

| | TS Source | Optimization method | Rules count |
|---|---|---|---|
| **FC1** | Math. model | Math. modelling | 125 |
| **FC2** | CO (GA-PID) | Math. modelling | 125 |
| **FC3** | Math. model | Remote connection | 125 |
| **FC4** | CO (GA-PID) | Remote connection | 125 |
| **FC5** | Math. model | Math. modeling + Remote connection | 125 |
| **FC6** | CO (GA-PID) | Remote connection + Math. modeling | 125 |

Experiment and modeling were performed in two control situations.

The first situation (S1) is typical for the control system (the initial angle equals to 1). The goal is to maintain the pendulum in equilibrium (0° angle of deflection). It should be noted that KB optimization held in this control situation.

The second situation is unexpected (S2). The initial angle equals to 5°. This situation characterizes the perturbation caused by external influences on CO.

*Figure 15* shows a comparison of integrals of squared error for all regarded regulators in a typical situation of control:

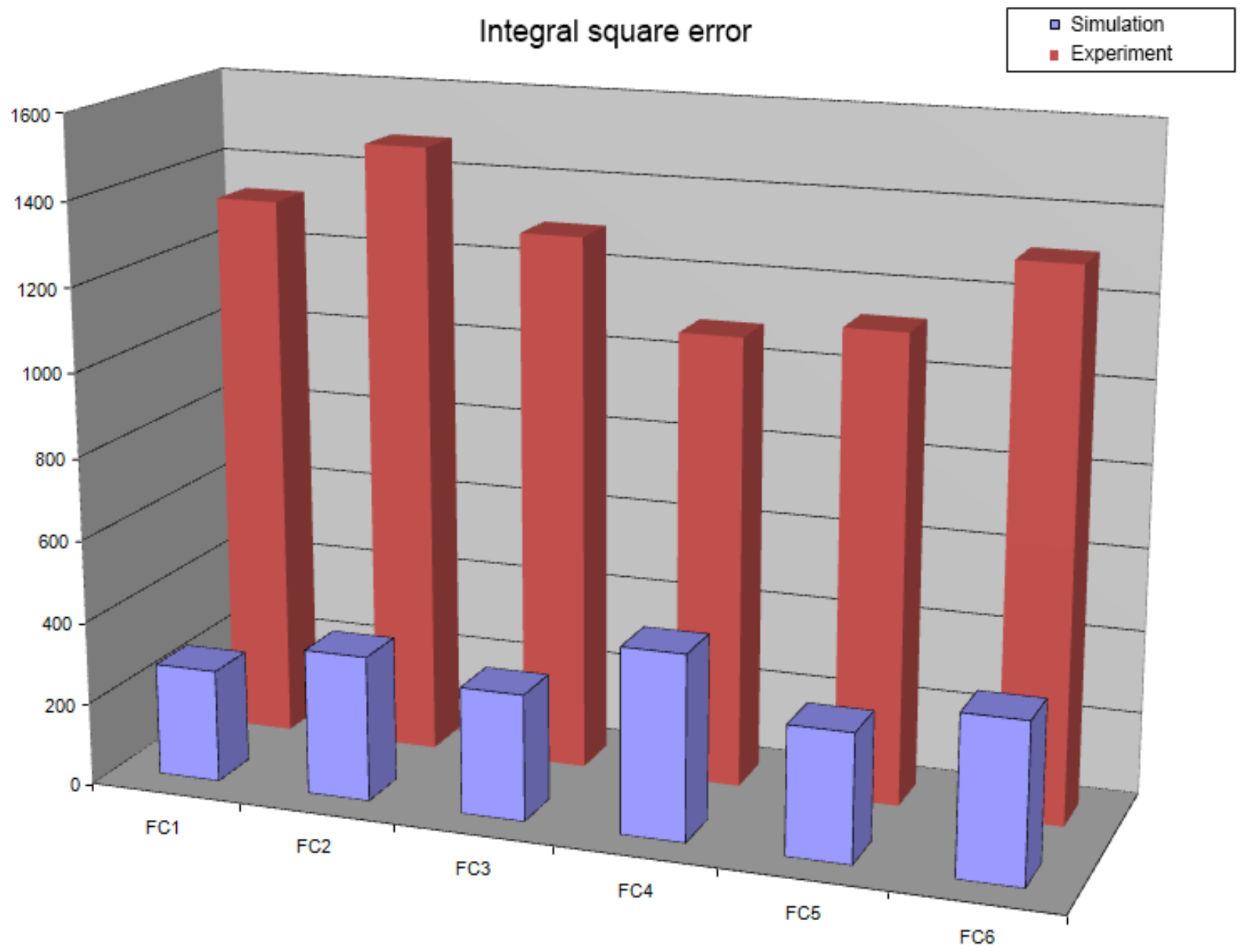

*Figure 15: Integral square error. Typical situation: Simulation and experiment.*

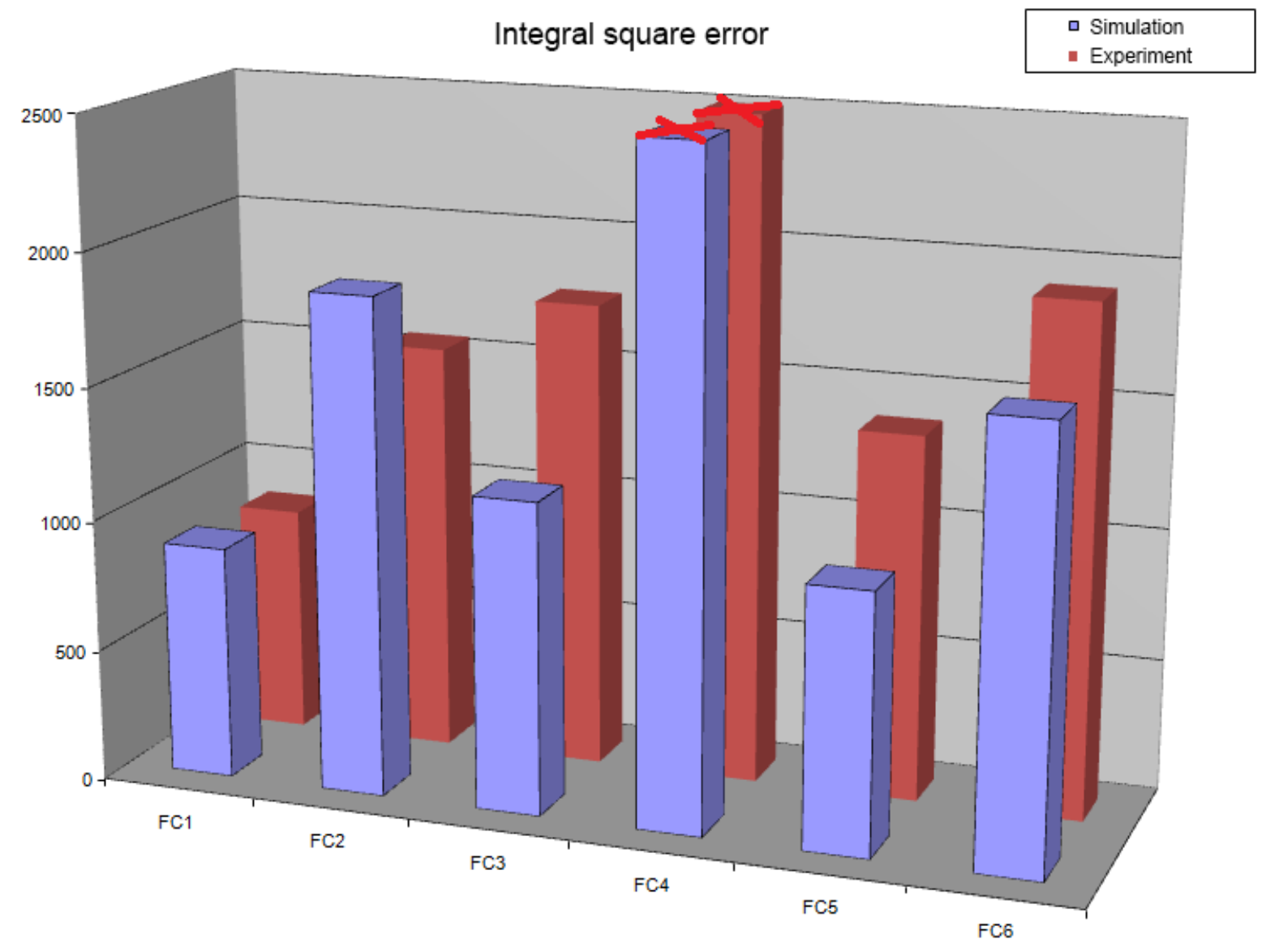

*Figure 16: Integral square error. Unpredicted situation: Simulation and experiment.*

The lower is integral square error level, the better controller works. Consider the results of simulation and experiment in unpredicted situation of control:

*Figure 16* shows a comparison of integrals of squared error for all regarded regulators in an unpredicted situation of control.

***L.*** *Benchmark 3: Remote quantum base optimization*

Let's compare the PID controller, fuzzy controllers $FC_1$ and $FC_4$, and QFI controllers based on different correlations: Quantum-Space (Q-S), Quantum-Time (Q-T), Quantum-Space-Time (Q-ST). These QFI controllers are optimized using remote connection.

Mathematical modeling and physical experiments took place in two control situation:

- in the first (typical) situation (S1), the delay of control is standard as 0.015 sec;
- in the second unpredicted situation (S2), the delay of the control as 0.035 sec.

From *Figs 17* and *18* it can be seen that KB optimization using a remote connection with quantum optimizer can improve the quality of control in a typical and unpredicted situation.

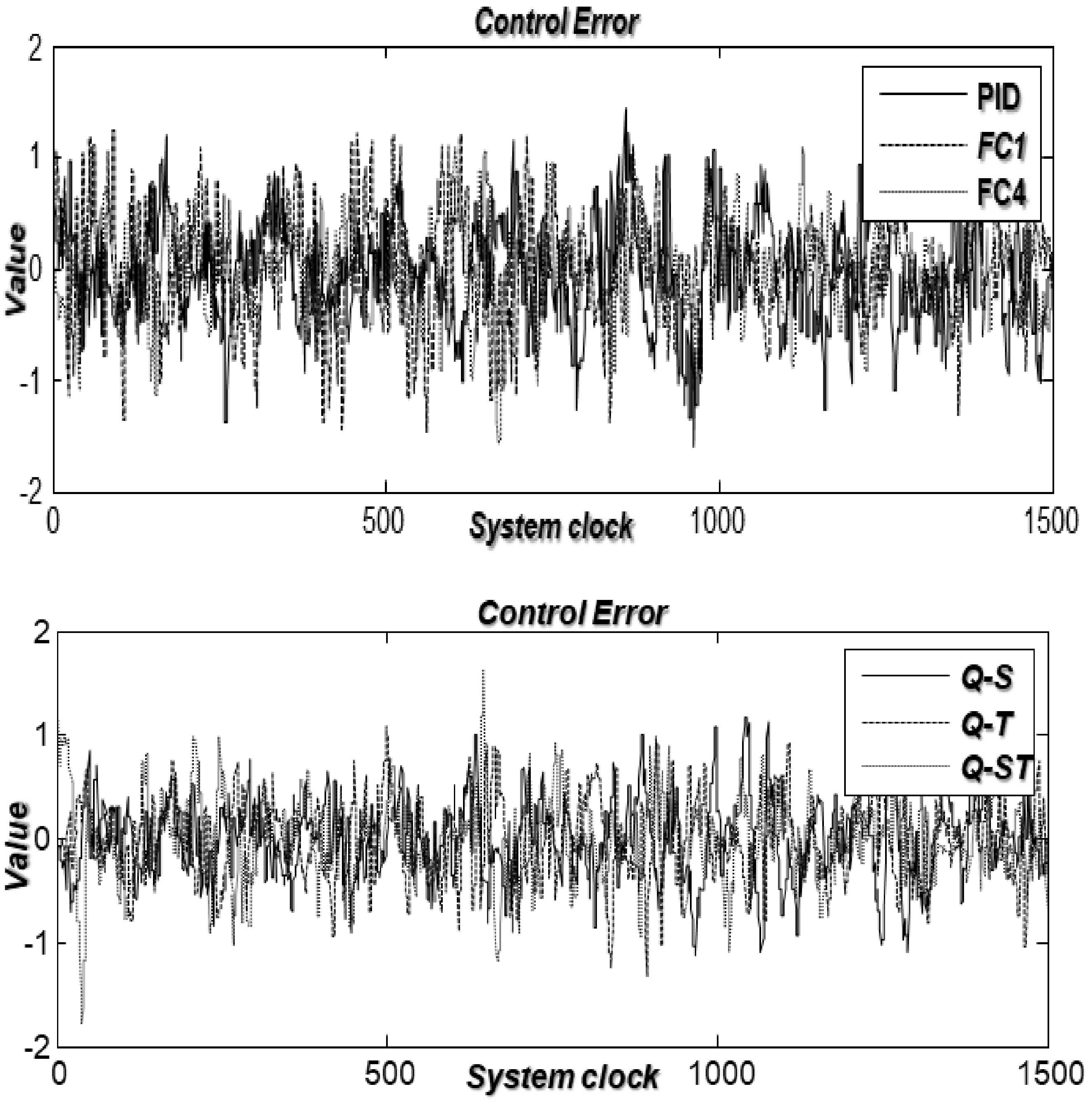

*Figure 17: Control error. Typical situation of control (Experiment).*

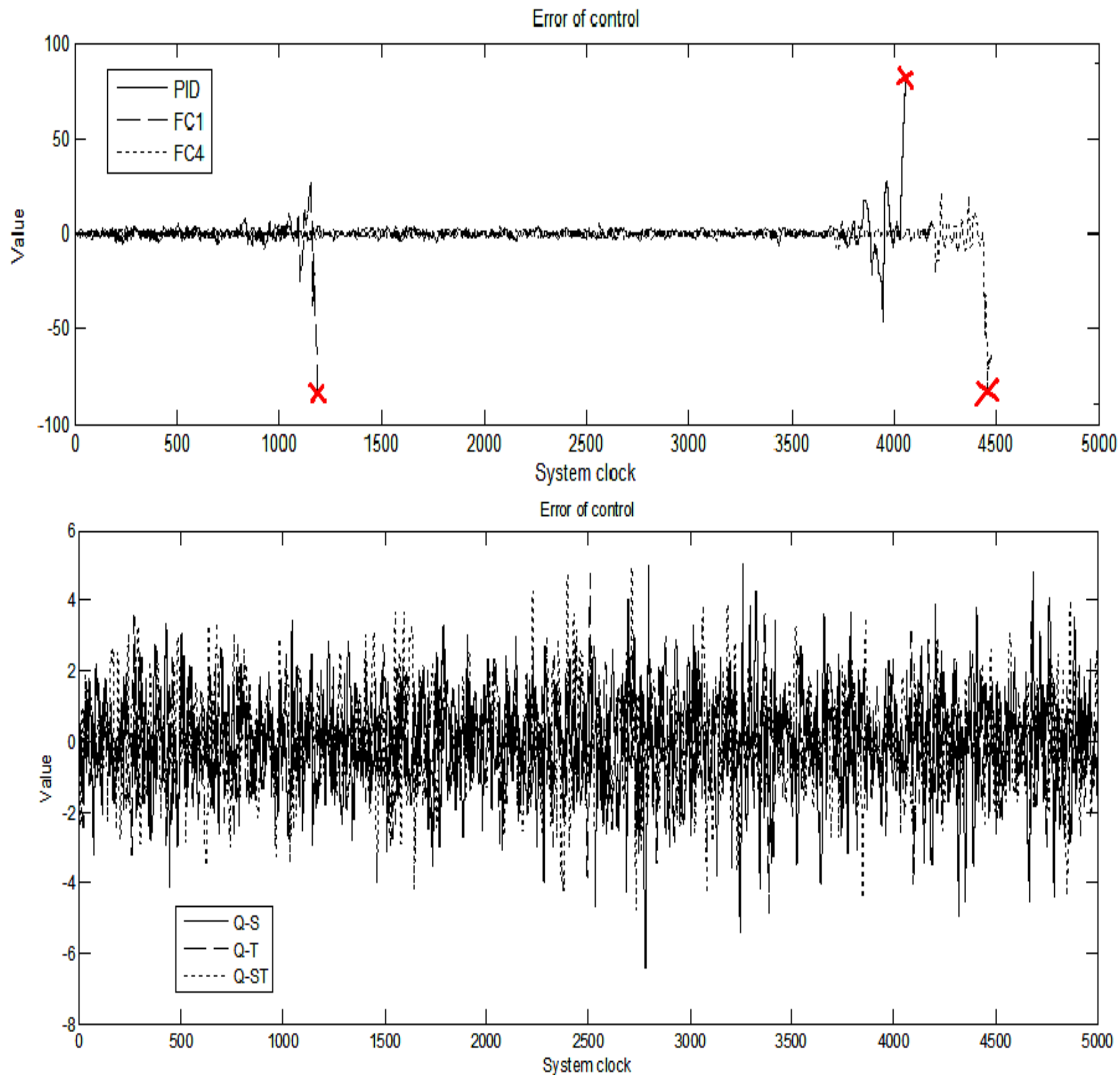

*Figure 18: Control error. Unpredicted situation of control (Experiment).*

*Related works.* Quantum computing approaching in robot path planning, emotion design, navigation, learning, decision making was applied also in [14-28] etc. Our approach is based on quantum self-organization of knowledge bases using responses of fuzzy controllers on unpredicted situations in on line.

## 6. Smart Robotic Manipulator: Quantum supremacy in intelligent control

The seven degrees of freedom (7 DoF) and seven-link robotic manipulator is described in this part. Due control object is complex, ICS for 7 DoF manipulator is constructed with using decomposition principal. Seven independent FCs (FC1 – FC7) are used for control each of manipulator link. The decomposition of control allows reducing complexity of constructing ICS. However, character of ICS somewhat reduced due to independence of seven FCs (*Fig. 19*).

QFI unit introduction allows improving ICS behavior by self-organization of independent KBs in FC1 – FC7. The correlation of three adjacent fuzzy KBs (the information FC $i$, $i+1$ and $i+2$ is used to control the $i$-th link of the manipulator, as shown on *Fig. 19(b)*. Consider the first internal unpredicted situation – the random noise in the control channel (see, the signal $s(t)$ on *Fig. 19*).

Comparison of manipulator behavior for control system based on soft computing and based on quantum soft computing in performance criteria terms is shown in *Fig. 20* (on the base results of sixty-five experiments).

The results are demonstrating if ICS is used with QFI gate (see, *Fig. 19 (a)*), all of evaluation of performance criteria improve (expect «One iteration time»).

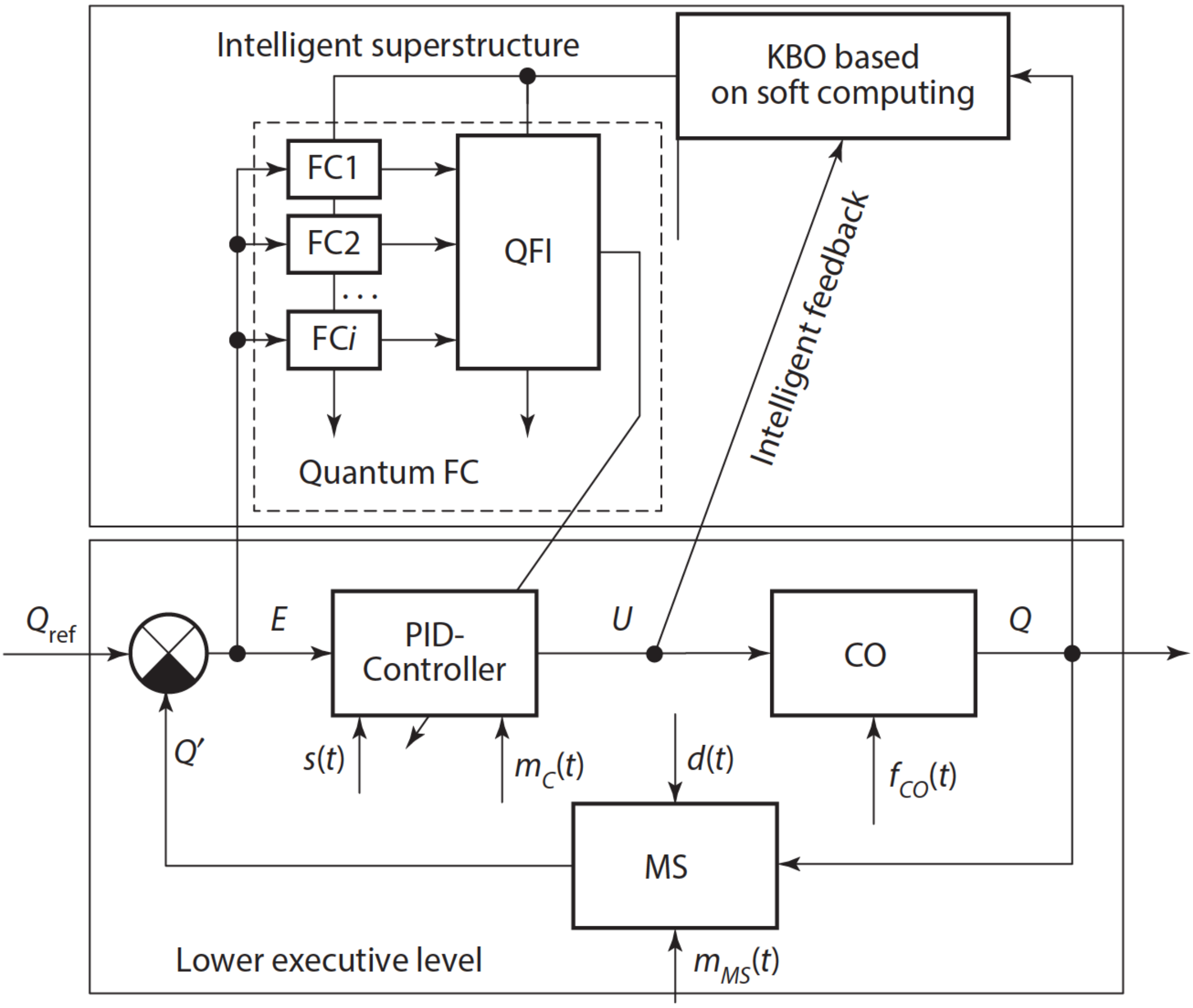

(a)

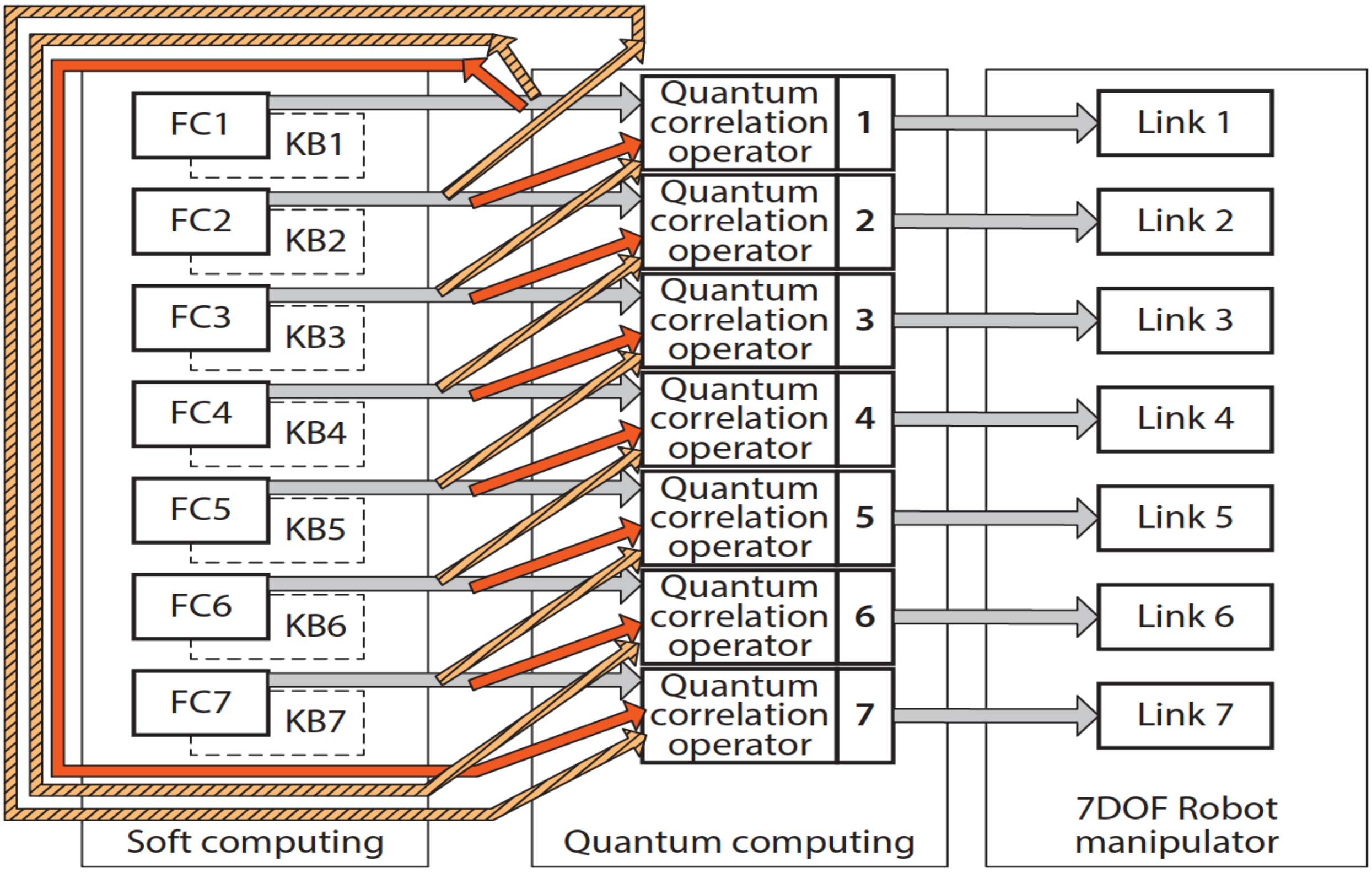

***(b)***

*Figure 19: (a) The structure of 7DoF manipulator ICS; (b) The application of the correlation of three neighboring FC.*

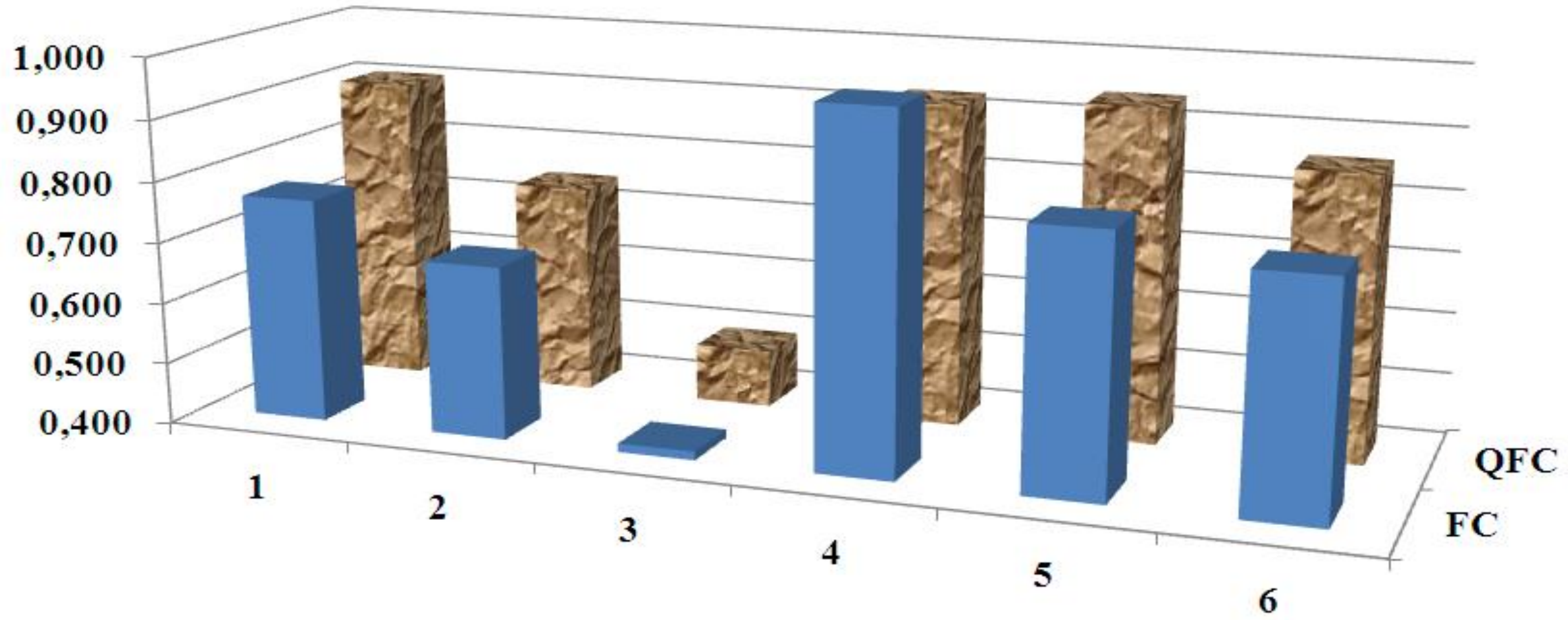

*Figure 20: Manipulator behavior with random noise in the control channel: FC – based on soft computing, QFC – based on quantum soft.*

The one of cases is shown in *Fig. 21 (a)*. Positioning accuracy is better if used ICS with QFI unit (in this case positioning error is 0.184 m). Positioning error is 1.918 m, if used ICS without QFI unit.

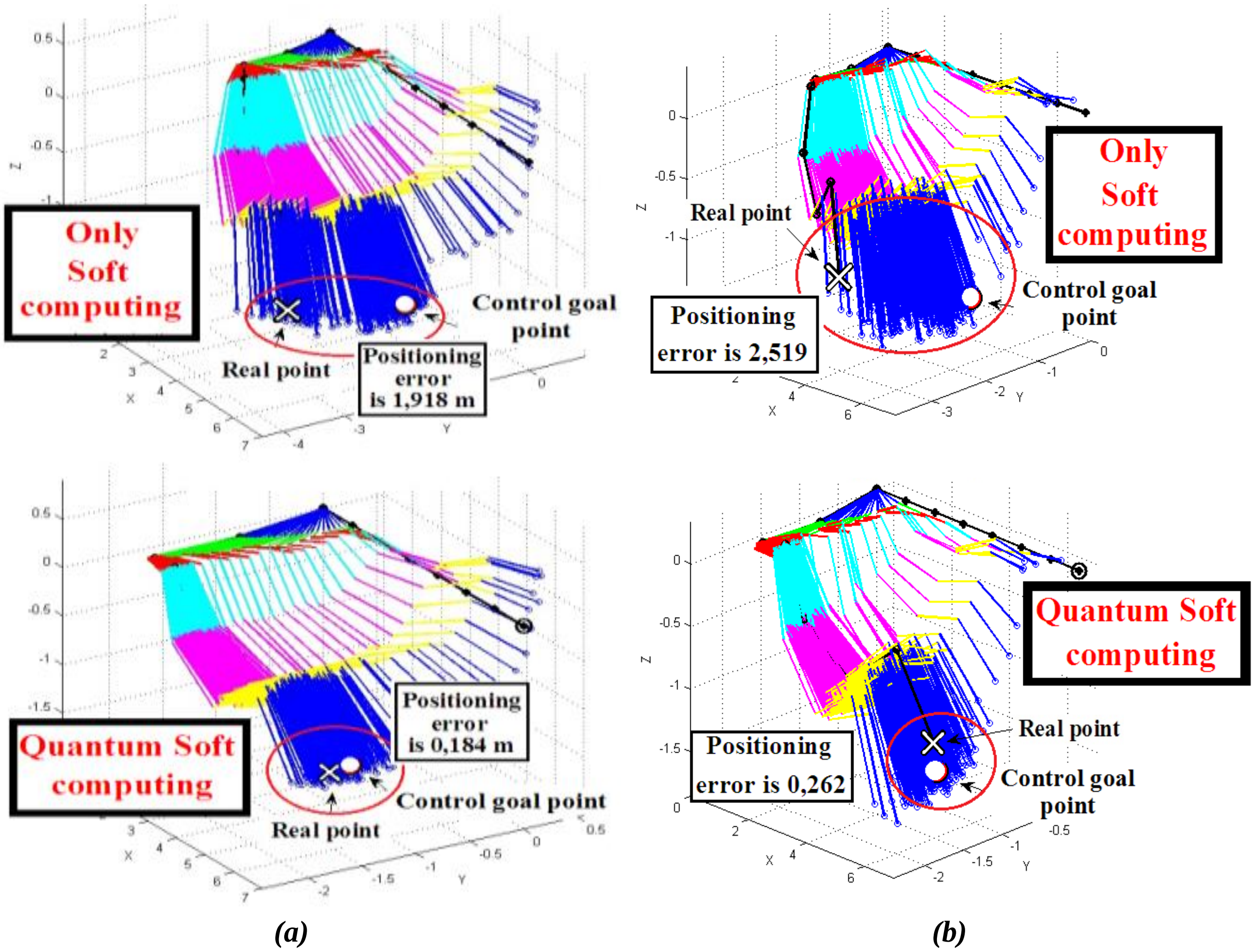

*Figure 21: (a) Manipulator behavior with random noise in the control channel, (b) Manipulator behavior with random noise in the measurement system.*

Consider the second internal unpredicted situation – random noise in the measurement system (see, the signal $d(t)$ and «Sensors», *Fig. 19(a)*). Comparison of manipulator behavior for control system based on soft computing and based on quantum soft computing in performance criteria terms is shown in *Fig. 22*.

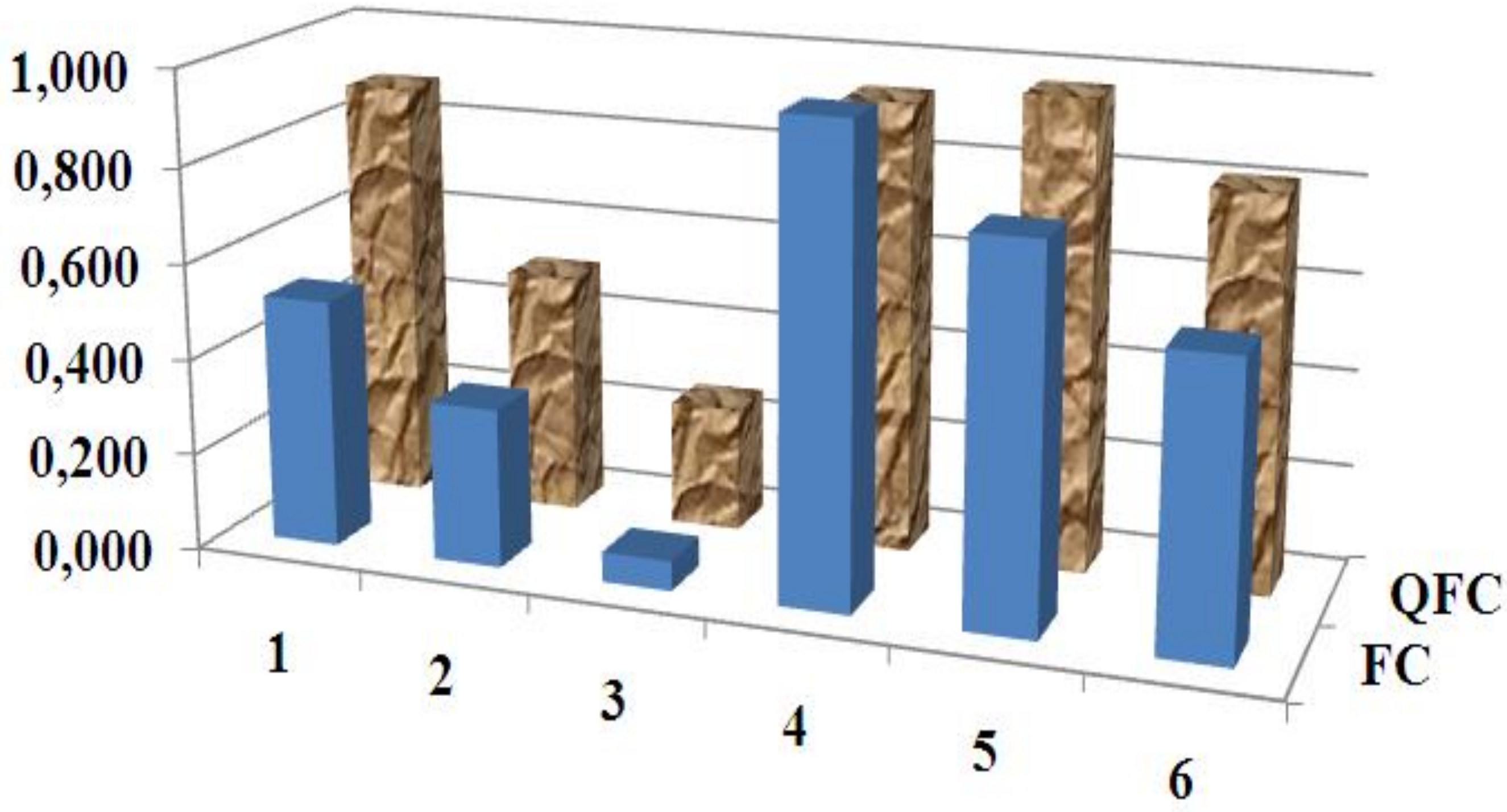

*Figure 22: Manipulator behavior with random noise in the measurement.*

The results are demonstrating if ICS is used with QFI unit, all of evaluation of performance criteria improve (expect «One iteration time»). The one of cases is shown in *Fig. 21 (b)*. Positioning accuracy is better if used ICS with QFI unit (in this case positioning error is 0.262 m). Positioning error is 2.519 m, if used ICS without QFI unit. Thus, the positioning accuracy ten times increased with QFI application in the comparison with the using case of soft computing and these facts demonstrate the quantum supremacy of described methods of robust control design [29,30].

## Conclusion

- New circuit implementation design method of quantum gates for fast classical efficient simulation of search QAs is developed. Benchmarks of design application as Grover's QSA and QFI based on QGA demonstrated.
- Applications of QAG approach in intelligent control systems with quantum self-organization of imperfect knowledge bases are described on concrete examples. Quantum supremacy on robotic Benchmarks demonstrated.
- Results of controller's behavior comparison confirm the existence of synergetic self-organization effect in the design process of robust KB on the base of imperfect (non-robust) KB of fuzzy controllers: from two imperfect KB with quantum approach a robust KB can be created using only quantum correlation. In classical intelligent control based on soft computing toolkit this effect impossible to achieve.
- Described approach opens new prospects for application of the model of quantum FC as the particular variant of the quantum self-organization algorithm in multi-objective control problems for the control object with weakly formalized structure and large dimensionality of the phase space of control parameters, application of experimental data in the form of the learning signal without development the mathematical model of the control object. These facts present a great advantage which is manifested as the possibility of design of control with required robustness in on line.

## Appendix. Maxwell's demon and quantum thermodynamic force

It was shown that a thermodynamic force is responsible for the flow and backflow of information in quantum processes [30-39]. For a system, interacting with a bath initially at temperature $\beta = 1/T$, the rate of the entropy production can be expressed as $\frac{d_i S}{dt} = Tr\left[F_{th} V_{th}\right]$, where $V_{th} \equiv \dot{\rho}_t \rho_t^{\beta}$ is the thermodynamic flow and $F_{th} \equiv \frac{1}{\rho_t^{\beta}}\left[\ln \rho_t^{\beta} - \ln \rho_t\right]$ the thermodynamic force. Then we get $\frac{dW_{irr}}{dt} = \frac{1}{\beta} Tr\left[F_{th} V_{th}\right]$. The thermodynamic force $F_{th}$ is responsible for the flow (encoding) and backflow (decoding) of information in Markovian and non-Markovian dynamics, respectively, Eqs suggests that, if the system is left to itself, $F_{th}$ actually encodes energy, during the flow, not to be used as work by the system and decodes energy, during the backflow, to be used as work by the system. In classical thermodynamics De Donder found a similar relation for chemical reactions. Let us now consider the case in which the system is not left to itself, i.e., someone or something outside the system (as a demon) intervenes in the process. Szilárd argued that negative work $\Delta W$ can be extracted from an isothermal cycle if Maxwell's demon plays the role of a feedback controller. When the statistical state of a system changes from $\rho(x)$ to $\rho(x|m)$ , due to the measurements made by the demon on the system, the change in the entropy of the system can be expressed as $\Delta S_{\text{meas}} = H\left(X|M\right) - H\left(X\right) = -I\left(X:M\right)$ , where $H\left(X\right) = -\sum_x \rho\left(x\right) \ln \rho\left(x\right)$ is the Shannon entropy of the system and $I\left(X:M\right)$ the mutual information between the state of the system and the measurement outcome $M$. Since $I\left(X:M\right)$ is always positive thus the demon causes the entropy of the system to decrease. This is similar to the case of non-Markovianity in which the entropy decreases. Therefore, the presence of the demon is also expected to lead to extracting more work from the system than what is expected. Now the role of the demon can be incorporated into the Second Law as $\Delta W \geq \Delta F - \frac{1}{\beta} I\left(X:M\right)$. Using a feedback controller (the demon) which makes measurements on the engine they are capable of extracting more work from the heat reservoirs than is otherwise possible in thermal equilibrium.

For a system, initially and finally in equilibrium states with temperature $\beta = 1/T$, which can contact heat reservoir $B_1, B_2, \ldots, B_n$ at respective temperatures $T_1, T_2, \ldots, T_n$ they have found that $\Delta W \geq \Delta F^{\beta} - \frac{1}{\beta} I\left(\rho_1 : X\right)$ and $I\left(\rho_1 : X\right) = \frac{1}{\beta}\left[S\left(\rho_1\right) - H\left(\left[p_k\right]\right) + H\left(\rho_1 : X\right)\right]$,
where $\rho_1$ is the state of the system at some time $t_1$, $S(\rho_1)$ the Von Neumann entropy, $H\left(p_k\right) = -\sum_k p_k \ln p_k$ the Shannon information content and

$$H\left(\rho_1 : X\right) = -\sum Tr\left(\sqrt{D_k}\rho_1\sqrt{D_k} \ln \sqrt{D_k}\rho_1\sqrt{D_k}\right).\ \left\{\sqrt{D_k}\right\}$$

are positive operator valued-measure (POVM) defined by $\sqrt{D_k} = M_k^{\dagger} M_k$ and $p_k = Tr\left(D_k \rho\right)$. It is seen that the sum of the last three terms on the right-hand side of the last inequality is the irreversible work due to the presence of the feedback controller (the demon). Thus, if we take the time derivative of these three terms, we have controller (the demon). Thus, if we take the time derivative of these three terms, we have

$$\frac{dW_{irr}^{dem}}{dt} = \frac{1}{\beta}\left[Tr\left(\dot{\rho}_1 \ln \rho_1\right) + \sum_k \dot{p}_k \ln p_k - \sum_k Tr\left(\sqrt{D_k}\dot{\rho}_1\sqrt{D_k} \ln \sqrt{D_k}\rho_1\sqrt{D_k}\right)\right].$$

It is observed that there are three quantum thermodynamic forces responsible for the extra work done during the process, $F_{th}^{1} = \frac{\ln \rho_1}{\rho_1^{\beta}}$, $F_{th}^{2(k)} = \frac{\ln p_k}{p_1^{\beta}}$, $F_{th}^{3(k)} = -\frac{\ln \sqrt{D_k}\rho_1\sqrt{D_k}}{\rho_1^{\beta}}$.

Thus, we may write $F_{th}^{tot} = F_{th}^1 \oplus F_{th}^2 \oplus F_{th}^3$. There are also *three thermodynamic flows associated with these three thermodynamic forces* above $V_{th}^1 = \dot{\rho}_1 \rho_1^\beta$, $V_{th}^{2(k)} = \dot{p}_k p_k^\beta$, $V_{th}^{3(k)} = \sqrt{D_k}\dot{\rho}_1\sqrt{D_k}\rho_1^\beta$ and it may be written $V_{th}^{tot} = V_{th}^1 \oplus V_{th}^2 \oplus V_{th}^3$.

These equations just indicate the fact that there are three thermodynamic forces and flows involved due to the presence of the feedback controller and we cannot add them up like the way we do about typical vectors [34,35]. We note that $F_{th}^{tot} = 0$ if and only if $D_k$ is proportional to the identity operator for all *k*, which means that nothing is intervening in the process, therefore no information is decoded to be used to perform additional work by the system. On the other hand, $F_{th}^{tot} = F_{th}^2$ if and only if $D_k$ is the projection operator satisfying $[\rho_1, D_k] = 0$ for all *k*, which means that the measurement on $\rho_1$ is classical, hence $F_{th}^{tot}$ is classical.

Therefore, it was shown that intervention (the demon) from the outside in the process of a system may be represented by a thermodynamic force.

***Example:*** *Quantum thermodynamic force and work extraction*. It is known that negative work $\Delta W$ can be extracted from an isothermal cycle with a feedback controller and due the measurement the change in the entropy of the system can be expressed as $\Delta S_{meas} = H(X|M) - H(X) = -I(X:M)$ and $H(X) = -\sum_x \rho(x)\ln\rho(x)$ is the Shannon entropy and $I(X:M)$ Is the mutual information between the state of the system and the measurement *M* outcome. Thus, while the amount of mutual information $I(X:M)$ is positive then the controller causes the entropy of the system to decrease. It is means that the presence of the controller expected to lead to extracting more work from the system that what expected. The controller can be incorporated into the Second Law as the following [36-40]

$$\Delta W \geq \Delta F^\beta - \frac{1}{\beta} I(\rho_1 : X) \text{ and } I(\rho_1 : X) = \frac{1}{\beta}\left[S(\rho_1) - H((p_k)) + H(\rho_1 : X)\right],$$

where $\rho_1$ is the state of the system at some time $t_1$, $S(\rho_1)$ is the von Neumann entropy, $H(p_k) = -\sum_k p_k \ln p_k$ is the Shannon entropy content and

$$H(\rho_1 : X) = -\sum_k Tr\left[\sqrt{D_k}\rho_1\sqrt{D_k}\ln\sqrt{D_k}\rho_1\sqrt{D_k}\right],\ (D_k) = M_k^\dagger M_k,\ p_k = Tr[D_k\rho]$$

are positive operator value-measure. The time derivative of these three terms give the result for extraction of work with controller as following:

$$\frac{dW_{irr}^{contr}}{dt} = \frac{1}{\beta}\left[Tr(\dot{\rho}_1 \ln \rho_1)\right] + \sum_k \dot{p}_k \ln p_k - \sum_k Tr\left[\sqrt{D_k}\dot{\rho}_1\sqrt{D_k}\ln\sqrt{D_k}\rho_1\sqrt{D_k}\right].$$

Thus, there are three quantum thermodynamic forces responsible for the extra work done during the process.

Since the work is given by the change in the internal, we can obtain

$$W = \Delta F(\alpha) + \alpha^{-1} D_{KL}\left[\rho_T \| \rho_{can,T}(\alpha)\right] - \alpha^{-1} D_{KL}\left[\rho_0 \| \rho_{can,0}(\alpha)\right]$$

where the conservation of the Gibbs-Shannon entropy in a thermally isolated Hamiltonian system apply, i.e., $S(\rho_T) = S(\rho_0)$ and $\Delta F(\alpha) \equiv F_T(\alpha) - F_0(\alpha)$ is the change in the Helmholtz free energy of the system. Since there is no heat bath and we do not have an a priori temperature, we call the parameter corresponding to an inverse temperature as *α*. The work equality is just a statement of energy accounting when we know the initial and final state of the system as well as the initial and final Hamiltonians. Under the condition of a known initial state the following work inequality can be achieved: $W \geq \Delta F(\alpha) - \alpha^{-1} D_{KL}\left[\rho_0 \| \rho_{can,0}(\alpha)\right] \equiv W_{LB}(\alpha)$, where the lower bound for the work denoted as $W_{LB}(\alpha)$. Thus, there exists a best value for α; namely, the value for which $W_{LB}(\alpha)$ is a maximum.